\documentclass[letterpaper,superscriptaddress,english,aps,prb,twocolumn,floatfix, showpacs, amsfonts, amssymb]{revtex4}
\usepackage{epsfig, graphicx,psfrag,amsmath,amssymb,float}
\usepackage[T1]{fontenc}
\usepackage{amsmath}
\usepackage{amssymb}

\usepackage{amscd}
\usepackage{bm}
\usepackage{psfrag}

\usepackage{babel}

\newcommand{\be}{\begin{equation}}
\newcommand{\ee}{\end{equation}}
\newcommand{\bea}{\begin{eqnarray}}
\newcommand{\eea}{\end{eqnarray}}
\newcommand{\ket}{\rangle}
\newcommand{\bra}{\langle}
\newcommand{\mc}{\mathcal}

\begin{document}

\title{Spin-charge coupling in quantum wires at zero magnetic field}

\author{Rodrigo G. Pereira}

\affiliation{Kavli Institute for Theoretical Physics, University of California,
Santa Barbara, CA 93106, USA }

\affiliation{Instituto de F\'{\i i}sica de S\~ao Carlos, Universidade de S\~ao Paulo, C.P. 369, 
S\~ao Carlos, SP.  13566-970, Brazil}

\author{Eran Sela}

\affiliation{Institute for Theoretical Physics, University of Cologne, 50937 Cologne, Germany}

\date{\today}

\begin{abstract}
We discuss  an approximation for the dynamic charge response of nonlinear spin-1/2 Luttinger liquids in the limit of small momentum. Besides accounting for the broadening of the charge peak due to two-holon excitations, the nonlinearity of the dispersion gives rise to a two-spinon peak, which at zero temperature has an asymmetric line shape. At finite temperature the spin peak is broadened by diffusion. As an application, we discuss the density and temperature dependence of the Coulomb drag resistivity due to long-wavelength scattering between quantum wires. 
\end{abstract}

\pacs{71.10.Pm, 73.63.Nm}

\maketitle

\section{Introduction}
There is theoretical consensus,\cite{tsvelik} supported by accumulating experimental evidence,\cite{auslaender,jompol} that in one dimension electrons decay into fractional excitations carrying either spin or charge, called spinon and holon. 
At low energies, interacting  one-dimensional (1D) systems  are described by the Luttinger model,\cite{tsvelik} which predicts that the collective spin and charge  modes  are decoupled and propagate with different velocities. Away from the low-energy limit, spin-charge separation holds in the sense that spinon and holon branches can still be identified in some momentum-resolved experiments such as angle-resolved photoemission,\cite{kim} but charge and spin degrees of freedom are inevitably  coupled by dispersion nonlinearity. A direct consequence is that at finite energies spin excitations can contribute to charge responses.\cite{nayak,dassarma}

Recently the effects of nonlinear dispersion in dynamical properties of Luttinger liquids have been emphasized.\cite{imambekov} In particular, the interplay of band curvature and interactions is essential for the interpretation of Coulomb drag experiments in parallel quantum wires.\cite{nazarov,debray} As discussed by Pustilnik {\it et al.},\cite{pustilnik} the Luttinger model cannot account for the leading contribution to the drag resistivity when there is a density mismatch between the wires, in which case  interwire backscattering processes are exponentially suppressed at low temperatures. The other type of low-energy process, long-wavelength scattering,  is ineffective within the Luttinger model because the dynamic charge structure factor (DCSF) $S(q,\omega)$ for small wavevector $q$  is given by a delta function peak at the energy of a free boson. In this approximation, the DCSFs of two wires with different densities have no overlap and the drag resistivity vanishes. For \emph{spinless} fermions,\cite{pustilnik} it is known that nonlinear dispersion is responsible for broadening the DCSF into a rectangular line shape with width proportional to $q^2$.\cite{rozhkov, pereiraJSTAT} This effect restores a smooth density dependence of the drag resistivity.

In this work we study the DCSF of spin-1/2 fermions in the limit of small $q$ and at zero magnetic field. Our motivation comes from the search for Luttinger liquid behavior in experiments  with  vertically coupled quantum wires, in which drag is enhanced by a  smaller interwire separation and the densities in each wire can be tuned independently.\cite{laroche} We are interested in the possibility that spinons  give a contribution to the drag resistivity  via spin-charge coupling at finite energies. This effect can not be described by the Luttinger model since it depends on violating particle-hole symmetry.    In order to calculate the DCSF, we follow the approach of treating band curvature as a perturbation to the Luttinger model, and resort to refermionization of the collective modes in the cases where perturbation theory is singular. While the drag response depends mostly on the spectral weight and width of the peaks as a function of wavevector and temperature, we also discuss other features of the DCSF that are of general interest for the dynamics of spin-1/2 fermions.   These features  could be directly probed by momentum-resolved techniques, such as   Bragg spectroscopy in cold Fermi gases.\cite{stoferle} We show that at zero temperature the charge peak has a $q^2$-scaling width, like in the spinless case, but  there is also a  peak due to spin excitations which resembles the dynamic spin structure factor (DSSF) of Heisenberg spin chains. At zero temperature, the DCSF diverges at the lower edge of the spin peak as a power law with exponent $\mu_{s-}=-1/2+\mc{O}(q^2)$. At finite temperature, the spin peak is broadened by diffusion.

The paper is organized as follows. In Sec. II we present the linear response formula for the drag resistivity and discuss its relation to the problem of the DCSF of spin-1/2 fermions with nonlinear dispersion. In Sec. III, we derive the effective bosonic Hamiltonian including irrelevant operators associated with band curvature, some of which couple charge and spin degrees of freedom. In Sec. IV, we describe the line shape of the DCSF in the limit of small $q$ at zero temperature. Finite temperature effects are also discussed. In Sec. V, the approximation for the DCSF is applied to calculate the density and temperature dependence of the drag resistivity. Finally, we summarize the results in Sec. VI. 

\section{Coulomb drag and dynamic charge response}

The drag resistivity between two capacitively coupled wires of length $L$ is defined as the ratio $r=-(e^2/2\pi\hbar)V_2/I_1L$, where $V_2$ is the voltage induced across  wire 2 (called the drag wire) when a current $I_1$ is driven through wire 1 (called the drive wire). For the typical setup, see Refs. \onlinecite{nazarov, debray}.  Let us assume clean wires ($L$ smaller than the mean free path due to impurities) and temperature regime $k_BT\gg \hbar v_{Fi}/L$, where $v_{Fi}$, $i=1,2$, is the Fermi velocity for electrons in each wire. The latter condition rules out finite size effects which are known to produce oscillations in the drag response as a function of drive voltage.\cite{peguiron} Hereafter we set $\hbar=k_B=1$. In the linear response regime, the drag resistivity at temperature $T$ is given by \cite{pustilnik}
\be
r=\frac{U^2}{4\pi^3 \nu_1 \nu_2T}\int_0^\infty dq\int_0^\infty d\omega\,\frac{q^2A_1(q,\omega)A_2(q,\omega)}{\sinh^2(\omega/2T)},\label{dragresist}
\ee
where $U$ is the interwire Coulomb interaction, $\nu_{i}$ is the charge density  and $A_{i}(q,\omega)$ is minus the imaginary part of the retarded  density-density correlation function in wire $i=1,2$. Eq. (\ref{dragresist}) expresses the drag resistivity as a functional of the dynamic density-density correlation function of two decoupled wires. Due to Boltzmann factors, the nonzero response comes from the overlap of $A_1$ and $A_2$  integrated up to frequencies of order $T$. At low temperatures compared to the Fermi energies $\epsilon_{Fi}$ of the wires, and neglecting interwire backscattering, the main contribution to the integral in Eq. (\ref{dragresist}) is due to small-$q$ (or forward) scattering.\cite{pustilnik} 

Our goal is to derive an aproximation for $A(q,\omega)$ in a single wire in the limit  $q\ll k_{Fi}$ and $\omega\ll \epsilon_{Fi}$. From this point until Sec. IV, we will be concerned with the dynamic response of a single wire and will omit the wire index $i=1,2$. The wire index will be restored in Sec. V when we return to Eq. (\ref{dragresist}) to compute the drag resistivity.

In order  to describe the intrawire interactions, we consider a Galilean-invariant model with  electron mass $m$ and  short-range density-density interaction potential $V(x)$
\begin{eqnarray}
 H&=&- \frac{1}{2 m} \int_0^L dx\,\Psi^\dagger \partial_x^2 \Psi \nonumber \\
& &+ \frac{1}{2} \int_0^L dx   \int_0^L dy \,V(x-y)n(x)n(y).\label{model}
\end{eqnarray}
Here $\Psi =(\psi_\uparrow,\psi_\downarrow)$ is a two-component fermionic field and  $n(x)=\Psi^\dagger(x) \Psi(x)$ is the local charge density. At zero magnetic field, the number of electrons with spin $\sigma=\uparrow,\downarrow$ is $N_\uparrow=N_\downarrow=N/2$. 
 The average density is  $\nu=N/L$ and the Fermi wavevector is $k_F=\pi \nu/2$.  We assume the interaction potential to have a finite range $R$, due to screening by nearby gates. For simplicity, we will discuss the properties of $A(q,\omega)$ in the thermodynamic limit.

The spectral function $A(q,\omega)$ in Eq. (\ref{dragresist}) satisfies the fluctuation-dissipation theorem \be2A(q,\omega)=(1-e^{-\omega/T})S(q,\omega),\ee where  $S(q,\omega)$ is the DCSF  given by \be
S(q,\omega)=\int_0^L dx\, e^{-iqx}\int_{-\infty}^{+\infty} dt\,  e^{i\omega t}\bra n(x,t) n(0,0)\ket.
\ee
 Since $S(-q,\omega)=S(q,\omega)$, hereafter we take $q>0$. Since we are interested in the regime $T\ll \epsilon_F$, we shall start by discussing the DCSF at $T=0$. The exact result for the noninteracting case, $V=0$, is \be
S_0(q,\omega)=(2m/q)\, \theta(q^2/2m-|\omega-v_Fq|),\label{Sfreefermion}
\ee
where $v_F=k_F/m$ is the Fermi velocity. For $V=0$, the spectral weight vanishes outside the  particle-hole continuum defined  by the upper and lower thresholds $\omega_\pm(q)=v_F q\pm q^2/2m$. As a function of energy $\omega$, the line shape of the DCSF in this case consists of one rectangular peak whose width is given by the band curvature scale $\delta\omega(q)=q^2/m$.

It is important to note that for spin-1/2 fermions the DCSF  at small $q$ cannot be obtained by perturbation theory in the interaction. In fact, the first order correction to $S(q,\omega)$  has logarithmic divergences at  $\omega\approx\omega_\pm(q)$\be
\frac{\delta S(q,\omega)}{S_0(q,v_Fq)}\approx \frac{m(2\tilde{V}_q-\tilde{V}_0)}{\pi q}\ln\left|\frac{\omega-\omega_-}{\omega_+ -\omega}\right|, \label{perturbation}
\ee
where $\tilde{V}_k$ is the Fourier transform of $V(x)$. These divergences signal edge singularities and are reminiscent of the result for spinless fermions,\cite{pustilnik06} but there is an important difference. For spinless fermions, the prefactor of the logarithmic divergence is $m(\tilde{V}_0-\tilde{V}_q)/\pi q$; for short-range interactions, $\tilde{V}_0 - \tilde{V}_q \sim q^2$, it vanishes as $q\to 0$. In contrast, the prefactor in the spinful case, $m(2\tilde{V}_q-\tilde{V}_0)/\pi q\approx m\tilde{V}_0/\pi q$,  diverges as $q\to 0$. The difference stems from the amplitude for $s$-wave scattering between electrons with opposite spin. This indicates that the limits $q\to 0$ and $\tilde{V}_0\to 0$ in $S(q,\omega)$ do not commute. In the limit $q\ll m\tilde{V}_0$, it is important to treat interactions exactly and account for the effects of spin-charge separation, as we will discuss in the following sections.

\section{Effective model for spin-charge coupling}
In the regime $q\ll m\tilde{V}_0$, we can treat interactions exactly and regard band curvature as a perturbation, with $q/k_F$ playing the role of a small parameter.\cite{pereiraJSTAT,teber,aristov} In the linear dispersion approximation, bosonization \cite{tsvelik} of Hamiltonian (\ref{model}) is standard and leads to the Luttinger model with Hamiltonian density\be
\mathcal{H}_\ell=2\pi v_c\left(J_R^2+J_L^2\right)+\frac{2\pi v_{s}}{3}\left(\mathbf{J}_{R}^{2}+\mathbf{J}_{L}^{2}\right)-2\pi v_{s}g\,\mathbf{J}_{R}\cdot\mathbf{J}_{L}.\label{eq:luttingermodel}\ee
Here $J_{R/L}$ ($\mathbf{J}_{R/L}$)  are  chiral  U(1) charge (SU(2) spin)  currents,  $v_c$ ($v_s$) is the charge (spin) velocity, and $g$ is the bare coupling constant of the marginally irrelevant backscattering operator. The long-wavelength part of the charge density fluctuation is \be n= 2\sqrt{K_c}(J_R+J_L),\label{nJRJL}\ee where $K_c$ is the Luttinger parameter for the charge sector.
 Galilean invariance implies $K_c=v_F/v_c$. At weak coupling, $\tilde{V}_0\ll v_F$, we have $v_c\approx  v_F+\tilde{V}_0/\pi$, $v_s\approx v_F$ and $g \approx \tilde{V}_{2k_F}/\pi v_F$.\cite{tsvelik}  In semiconductor quantum wires, typical values of  $K_c\approx 0.7$ have been reported.\cite{jompol} 
 
 The charge and spin currents can be expressed in terms of chiral bosonic fields as \bea J_{R/L}&=&\mp \partial_x\varphi^c_{R/L}/\sqrt{4\pi}, \\ J^z_{R/L}&=&\mp \partial_x\varphi^s_{R/L}/\sqrt{4\pi},\eea which obey the commutation relations $[\varphi^c_{R/L}(x),\partial_{x^\prime}\varphi^c_{R/L}(x^\prime)]=[\varphi^s_{R/L}(x),\partial_{x^\prime}\varphi^s_{R/L}(x^\prime)]=\mp i\delta(x-x^\prime)$. The transverse part of the spin currents can be written as \bea J_{R/L}^+ &=& J_{R/L}^x + i J_{R/L}^y=\frac{1}{2 \pi}\, e^{ +i \sqrt{4 \pi} \varphi^s_{R/L}}, \\
 J_{R/L}^- &=& J_{R/L}^x - i J_{R/L}^y=\frac{1}{2 \pi}\, e^{- i \sqrt{4 \pi} \varphi^s_{R/L}},
 \eea
where the short-distance cutoff is set to 1.

The leading (dimension-three) perturbations to model (\ref{eq:luttingermodel}), generated by the quadratic term in the electron dispersion as well as irrelevant interactions, are\bea
\delta \mathcal{H} &=& (4\pi^2/3)\left[\eta_-(J_R^3+J_L^3)-\eta_+(J_R^2J_L+J_L^2J_R)\right.\nonumber\\
&&\left.+\kappa_- (J_R\mathbf{J}_R^2+J_L\mathbf{J}_L^2)+\kappa_+ (J_R\mathbf{J}_L^2+J_L\mathbf{J}_R^2)\right.\nonumber\\
&&\left.+\kappa_3 (J_L+J_R) \mathbf{J}_L \cdot \mathbf{J}_R\right].\label{H3}
\eea
The last three terms in Eq. (\ref{H3}) couple spin and charge. Importantly,  only even powers of the spin currents are allowed in $\delta\mc{H}$ due to SU(2) symmetry. Direct bosonization of Hamiltonian (\ref{model}) produces all terms in Eq. (\ref{H3}), except for the $\kappa_3$ term. This does not mean that $\kappa_3$ vanishes (such term is allowed by symmetry), but rather that it must be generated at second order in the electron-electron interaction.  

In fact, we can derive phenomenological relations for all coupling constants. The exact parameters $\eta_\pm$  can be related to the change of $v_c$ and $K_c$ under a shift of chemical potential $\mu$. The calculation is analogous to the spinless case in  Ref. \onlinecite{pereiraJSTAT}; simplifying for the case of Galilean invariance where $K_c=v_F/v_c$ and using the result for the compressibility $\partial \nu /\partial \mu=2K_c/\pi v_c$, we find \bea
\eta_-&=&\frac{1}{2\sqrt{K_c}}\left(\frac1m+v_c\frac{\partial v_c}{\partial \mu}\right),\label{etam}\\
\eta_+&=&\frac{3}{2\sqrt{K_c}}\left(\frac1{m}-v_c\frac{\partial v_c}{\partial \mu}\right).
\eea

Likewise, an infinitesimal chemical potential shift $\delta\mu$ modifies the spin velocity $v_s$ by giving a finite expectation value to the charge currents $\langle J_L\rangle=\langle J_R\rangle=\delta\mu \sqrt{K_c}/(2\pi v_c)$ in  the $\kappa_\pm$ terms in Eq. (\ref{H3}):\be
\frac{4\pi^2}{3}(\kappa_- +\kappa_+)\langle J_R\rangle (\mathbf{J}_R^2+\mathbf{J}_L^2)\equiv \frac{2\pi}{3}\delta v_s (\mathbf{J}_R^2+\mathbf{J}_L^2).\ee This relation  fixes the sum \be
\kappa_-+\kappa_+=(v_c/\sqrt{K_c})\,\partial v_s/\partial\mu.\label{sumkappas}\ee
Moreover, Galilean invariance requires that the charge current and momentum operators for model (\ref{eq:luttingermodel}) plus (\ref{H3}) be proportional to each other.\cite{nayak} The momentum operator is obtained from the energy-momentum tensor; its density is\be
\mathcal{P}=\frac{2k_F}{ \sqrt{K_c}}(J_R-J_L)+2\pi (J_R^2-J_L^2)+\frac{2\pi}{3}(\mathbf{J}_R^2-\mathbf{J}_L^2).
\ee
The current density $\mathcal{J}(x)$ is obtained from the continuity equation for the charge density:\be
\partial_t n(x)=-i\int  dx^\prime\,[n(x),\mathcal{H}_\ell(x^\prime)+\delta \mathcal{H}(x^\prime)]=-\partial_x \mathcal{J}(x).
\ee
We then impose the condition $\mathcal{J}(x) =\mathcal{P}(x)/m$ for a Galilean-invariant system. The relation between the coefficients of the spin contributions to $\mathcal{P}(x)$ and $\mathcal{J}(x)$  leads to \be
\kappa_--\kappa_+=1/(\sqrt{K_c}m).\label{diffkappas}
\ee
Eqs. (\ref{sumkappas}) and (\ref{diffkappas}) allow one to determine $\kappa_\pm$ by simply measuring the spin dispersion at low energies. 

Finally, the coefficient $\kappa_3$ is related to the variation of the backscattering coupling constant $v_s g$ under a change of the chemical potential,
\be
\label{kappa3}
\kappa_3 = -\frac{3 v_c}{2 \sqrt{K_c}} \frac{\partial (v_s g)}{\partial \mu}.
\ee
Since $g$ is marginally irrelevant, we expect $\kappa_3$ to be more irrelevant than the other coupling constants in Eq. (\ref{H3}), in the sense of logarithmic corrections to scaling. This will be discussed in the following subsection.

\subsection{Renormalization group flow with irrelevant operators}
\label{se:RG}
All the operators in Eq.~(\ref{H3}) are irrelevant and have the same scaling dimension $x=3$. The  renormalization group (RG) equations for the irrelevant coupling constants (including the marginal $g$ in Eq. (\ref{eq:luttingermodel})) can be derived by integrating out high-energy modes in the partition function as one lowers the ultraviolet momentum cutoff  $\Lambda$. Following the notation of Ref.~\onlinecite{cardy}, we define ``dimensionless'' coupling constants (which have dimensions of velocity) $\tilde{\eta}_\pm =\Lambda \eta_\pm $, $\tilde{\kappa}_\pm =\Lambda \kappa_\pm $ and $\tilde{\kappa}_3 =\Lambda \kappa_3 $. To obtain the quantum corrections to scaling, we use the operator product expansion (OPE) of the spin currents\cite{tsvelik} 
\begin{eqnarray}
\label{OPE0}
:J_{L}^a(z):\,: J_{L}^b(0): &\sim &\frac{\delta^{ab}}{8 \pi^2z^2} + \frac{i \varepsilon^{a b c}}{2 \pi z} :J^c_L(0):,\nonumber\\
:J_{R}^a(\bar z):\, :J_{R}^b(0): &\sim &\frac{\delta^{ab}}{8 \pi^2\bar z^2} + \frac{i \varepsilon^{a b c}}{2 \pi \bar z} :J^c_R(0):,\label{OPE}
\end{eqnarray}
where $z=v_s\tau+ix$ and $\bar z=v_s\tau-ix$, with $\tau$ the imaginary time, and $\varepsilon^{abc}$ is the Levi-Civita antisymmetric tensor.  [The normal ordering symbol $::$ is implicit in the Hamiltonian Eqs.~(\ref{eq:luttingermodel}) and (\ref{H3}).]   The OPE for the charge fields is simply
\begin{eqnarray}
:J_{L}(x,\tau):\,: J_{L}(0,0): &\sim &\frac{1}{8 \pi^2(v_c\tau+ix)^2}+\dots ,\nonumber\\
:J_{R}(x,\tau):\, :J_{R}(0,0): &\sim &\frac{1}{8 \pi^2(v_c\tau-ix)^2} +\dots.\label{OPEcharge}
\end{eqnarray}

We integrate out high-energy modes in the shell $1/\Lambda<|z|<1/\Lambda^\prime$ with $\Lambda^\prime=\Lambda\, e^{-d\ell}$, $d\ell \ll 1$. This choice of cutoff is rotationally invariant for the spin modes, but elliptical for the charge modes. In order to get a nonzero contribution to the RG equation after integrating out the shell in the $(x,\tau)$ plane, it is important to contract both right and left movers for a given species (charge or spin) at the same time.

In the presence of the dimension-three operators, the velocities, Luttinger parameter and chemical potential are renormalized, but flow to their fixed-point values in the low-energy limit. This flow is already taken into account if we use the exact parameters. The interesting RG flow here is given by the coupled equations for $\tilde{\kappa}_3$ and $g$\bea
\frac{dg}{d\ell}&=&-g^2,\label{RGg}\\
\frac{d\tilde\kappa_3}{d\ell}&=&-(1+2g)\tilde\kappa_3.\label{RGzeta}
\eea
There are no corrections to the scaling of $\tilde\eta_\pm$ and $\tilde\kappa_\pm$ to second order in the coupling constants. Eq. (\ref{RGzeta}) can be rewritten as\be
\frac{d\kappa_3}{d\ell}=-2g\kappa_3.\label{RGzeta2}\ee
On the right-hand side of Eqs. (\ref{RGg}) and (\ref{RGzeta2}) we have terms of zeroth order in $\Lambda$. It follows that\be
\frac{d\ln g}{d\ell}=-g=\frac12\frac{d\ln \kappa_3}{d\ell}.
\ee
The solution is of the form\be
\kappa_3(\ell)/[g(\ell)]^2=\textrm{ const.}.\label{solRG}
\ee
The scaling of the marginal coupling constant is the familiar one\be
g(\Lambda)=\frac{g}{1+g\ln(\Lambda_0/\Lambda)},\label{gscaling}
\ee 
where $\Lambda_0$ is the initial value of the cutoff.
As a result,  for positive $g\ll1$ and for $\Lambda\ll\Lambda_0e^{-1/g}$ the effective $g(\Lambda)$ vanishes logarithmically as $g(\Lambda)\sim 1/\ln(\Lambda_0/\Lambda)$. If the bare $g$ at high energies is of order 1, the perturbative  result in  Eq.  (\ref{gscaling}) is still valid if $g$ is interpreted as $g(\Lambda_0)$ at some scale $\Lambda_0\ll k_F$ such that $g(\Lambda_0)\ll 1$. In any case, we obtain $g(\Lambda)\sim 1/\ln(\Lambda_0/\Lambda)$ in the low-energy limit. 

More interestingly, Eq. (\ref{solRG}) implies \be
\kappa_3(\Lambda)=\frac{\kappa_3}{[1+g\ln(\Lambda_0/\Lambda)]^2}.\label{zetaln2}
\ee
Therefore, $\kappa_3(\Lambda)$ vanishes as $\kappa_3(\Lambda)\sim1/\ln^2(\Lambda_0/\Lambda)$ as $\Lambda\to0$. This will be important in Section \ref{sec:smear} when we compare leading logarithmic corrections to the DCSF due to $g$ and $\kappa_3$.

\section{Dynamical charge structure factor} 
In the bosonized form of Eq. (\ref{nJRJL}), the DCSF is given by 
\be
S(q, \omega) =  - 8 K_c {\rm{Im}}~C^{\textrm{ret}}(q ,  \omega) , \ee
where $C^{\textrm{ret}}(q ,  \omega) $ is the retarded  correlation function for the charge current $J_R+J_L$, which can be obtained by analytic continuation from the Matsubara correlation function\bea
C(q, i \omega) &=&\sum_{\alpha,\beta=R/L} C_{\alpha \beta}(q, i \omega) \\
C_{\alpha \beta}(q, i \omega) &=& - \int_0^L dx \,e^{-i q x} \int_0^\beta d \tau \,e^{i \omega \tau} \nonumber \\
&&  \times\langle J_{\alpha}(x,\tau)  J_{\beta}(0,0) \rangle.\label{chargeprop}
\eea
Eq. (\ref{chargeprop}) involves the charge boson propagator. Within the Luttinger model, the charge boson is free and we have $C_{LL}^{0}=C_L$,  $C_{RR}^{0}=C_R$ and $C_{LR}^0=C_{RL}^0=0$, with 
\begin{eqnarray}
C_{R/L}(x,\tau)&\equiv&\langle J_{R/L}(x,\tau) J_{R/L}(0,0)\rangle \nonumber\\
&=& \frac{1}{8 \pi^2} \frac{1}{(v_c \tau \mp i x)^2}.
\end{eqnarray}
Taking the Fourier transform, we obtain 
\begin{eqnarray}\label{eq:G0}
C_{R/L}(q ,i \omega) =\frac{1}{4 \pi}\frac{\pm q}{i \omega \mp v_cq }.
\end{eqnarray} 
As a result,  the DCSF calculated in the linear dispersion approximation is given by\be
S(q,\omega) = 2K_cq\,\delta(\omega-v_cq).\ee 
That the DCSF is given by a delta-function peak at the energy of the free charge boson follows from spin-charge separation and Lorentz invariance of the Luttinger model. This should be contrasted with  the free-electron result in Eq. (\ref{Sfreefermion}), where the peak associated with  particle-hole excitations has  a $q^2$ broadening due to the curvature of the dispersion about the Fermi points.

\subsection{Width of the charge peak\label{sec:chargepeak}}

We can calculate  $S(q,\omega)$ beyond the Luttinger liquid result by analyzing the effects of the boson-boson interactions in Eq. (\ref{H3}).  First, let us consider the broadening of the charge peak. The charge-only $\eta_\pm$ terms are  familiar from the spinless case.\cite{pereiraJSTAT} They account for the decay of one charge boson into two charge bosons. In particular, $\eta_-$ is a three-leg vertex in which one right- (left-) moving boson decays into two right-(left-)moving bosons, thus coupling the single-boson state to degenerate multi-boson states. It is known that perturbation theory in $\eta_-$ is badly divergent, but can be dealt with by refermionization.\cite{imambekov,rozhkov}  Near the charge mass shell, $\omega\approx v_cq$, we introduce a spinless \emph{holon} field $\psi_{c,R}$ such that $\psi^\dagger_{c,R}\psi^{\phantom\dagger}_{c,R}=\sqrt{2}J_{R}$. The $\eta_-$ term in Eq. (\ref{H3}) then maps onto a parabolic dispersion about the holon Fermi point \be
\frac{4\pi^2}{3}\eta_-J_R^3\to-\frac{\eta_-}{2\sqrt{2}}\psi^\dagger_{c,R}\partial_x^2\psi^{\phantom\dagger}_{c,R}.\ee 
It can be argued that $\eta_-$ determines the exact broadening of the DCSF  to order $q^2$ because it gives rise to the   bosonic diagrams that are most singular at $\omega= v_c q$.\cite{pereiraJSTAT}
Within the approximation of neglecting the other dimension-three operators, the charge sector of the Luttinger model plus the $\eta_-$ term refermionizes into a free fermion model with dispersion $\epsilon_c(k)\approx v_c k+\eta_-k^2/(2\sqrt{2})$, for $k$ measured from the right Fermi point. The support of the charge peak in the DCSF is then given by the spectrum of excitations with a single holon-anti-holon pair. Due to the curvature of the dispersion, these excitations define a continuum bounded by \be\omega_{c\pm}(q)=v_cq\pm \eta_-q^2/2\sqrt{2}.\label{chargewidth}\ee 
Therefore, at order $q^2$, the charge peak has a free-fermion-like line shape \be
S(q,\omega)\approx \frac{2\sqrt{2}K_c}{\eta_-q}\theta\left(\frac{\eta_-q^2}{2\sqrt{2}}-|\omega-v_cq|\right).\label{chargepeak}
\ee
The parameter $\sqrt{2}/\eta_-$ can be interpreted as a renormalized holon mass.

It is interesting that the limits  $q\to 0$ and $\tilde V_0\to0$ in the width  do not commute. For $q\ll m\tilde V_0\ll k_F$, we have from Eq. (\ref{etam}) that  $\delta\omega_c(q)=\eta_-q^2/\sqrt{2}\approx q^2/(\sqrt{2}m)$. The $\sqrt{2}$ factor makes the charge peak narrower than the free electron result in Eq. (\ref{Sfreefermion}).\cite{comment1} In particular, this means that in the regime $q\ll m\tilde{V}_0$ the holon dispersion (which shows up, for instance, in the single-electron spectral function) should not be regarded as a smooth continuation of the electron dispersion. An important crossover happens at $q\sim m\tilde{V}_0$.

The result in Eq. (\ref{chargewidth}) can be directly compared with the exact width of the two-holon continuum for an integrable model, such as the Yang-Gaudin model.\cite{yanggaudin} We have numerically solved the standard Bethe ansatz integral equations for the spectrum of elementary excitations of the Yang-Gaudin model.  We verified  that the width defined as the difference between the maximum and minimum exact energies of a particle-hole excitation in the holon Fermi sea for momentum $q\ll m(v_c-v_s)$ agrees with Eq. (\ref{chargewidth}), including the factor of $\sqrt2$ and with $\eta_-$ calculated from the phenomenological relation in Eq. (\ref{etam}).

Corrections to Eq. (\ref{chargepeak}) due to residual holon-holon interactions are higher order in $q$. These corrections include  a high frequency tail at order $\eta_+^2$, analogous to the spinless case,\cite{pustilnik,pereiraJSTAT,teber} and possible asymmetries of the charge peak near the edges of the two-holon continuum, due to x-ray edge type singularities.\cite{pustilnik06} But before we discuss the behavior near  $\omega_{c\pm}(q)$, we turn to the contributions from the spin operators in Eq. ($\ref{H3}$).

\subsection{Spin peak}

\label{se:spinpeak}
\begin{figure}
\begin{center}
\includegraphics*[width=70mm]{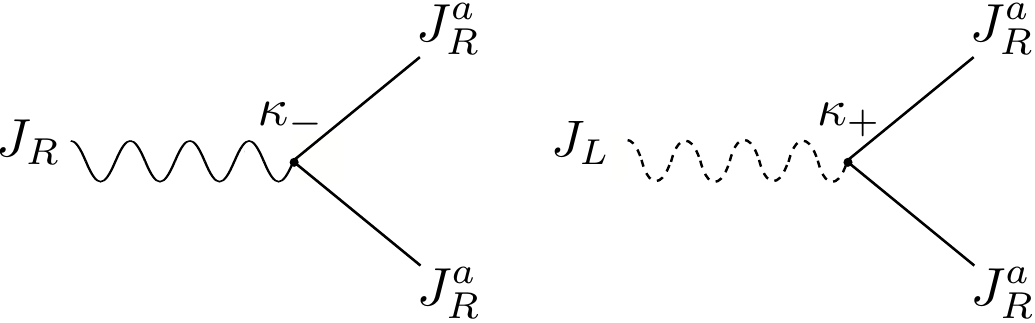}
\caption{Decay of a charge boson (propagators denoted by wiggly lines) into two right-moving spin bosons (propagators denoted by straight lines). This process leads to a spin peak in the dynamic charge structure factor.   \label{fg:diagram1}}
\end{center}
\end{figure}

The $\kappa_{\pm}$ operators in Eq. (\ref{H3}) allow for decay of the charge boson into  two spin bosons moving in the same direction and carrying the total energy $\omega\approx v_sq$. The corresponding three-leg vertices are illustrated in   Fig. \ref{fg:diagram1}. 
As noted in Ref. \onlinecite{teber}, this process gives rise to a narrow peak in the DCSF centered at $\omega=v_sq$ which corresponds to a charge-carrying spin singlet excitation. Let us calculate the correction to the charge boson propagator in Eq. (\ref{chargeprop}) using second order perturbation theory in $\kappa_\pm$. For instance, the $\mathcal{O}(\kappa_-^2)$ correction to $C_{RR}$ is
\begin{eqnarray}
\delta C_{RR}^{\kappa_-}(q,i \omega) =32 \pi^4  \kappa_-^2 [C_{R}(q, i \omega)]^2 \Pi_{RR}(q,i \omega),\label{piRR}
\end{eqnarray}
where we made use of the identity $:\mathbf{J}_{R/L}^2 := 3 (J^z_{R/L})^2$ and introduced the boson self-energy
\begin{eqnarray}
\label{eq:Pi}
\Pi_{\alpha \beta}(q,i \omega) &=& - \int dx\, e^{-i q x} \nonumber\\&&\times\int_0^\beta d \tau e^{i \omega \tau}\,S_\alpha(x,\tau)S_\beta(x,\tau).
\end{eqnarray}
Here $S_\alpha$, $\alpha=R,L$, are the free chiral spin boson propagators\bea
\delta^{ab}S_{R/L}(x,\tau)&=&\langle J^a_{R/L}(x,\tau)  J^b_{R/L}(0,0) \rangle\nonumber\\
&=&\frac{\delta^{ab}}{8 \pi^2} \frac{1}{(v_s \tau \mp i x)^2}.
\eea
In momentum and frequency space, we have
\begin{eqnarray}\label{eq:G0s}
S_{R/L}(q ,i \omega) = \frac{1}{4 \pi}\frac{\pm q}{i \omega \mp v_s q}. 
\end{eqnarray}We  then calculate $\Pi_{RR}$ that appears in Eq. (\ref{piRR}) 
\begin{eqnarray}
(4 \pi)^2 \Pi_{RR}(q,i \omega) &=&- \int_{0}^\infty \frac{dq'}{2 \pi} \int_{-\infty}^\infty \frac{d \omega'}{2 \pi} \frac{q'}{i \omega' -v_sq' } \nonumber\\&&\times \frac{q- q'}{i \omega- i \omega' -v_s q +v_sq'}\nonumber \\
&=&\frac{q^3}{12 \pi} \frac{1}{i \omega -v_sq }.
\end{eqnarray}
Adding up all second order contributions from $\kappa_-$ and $\kappa_+$ and taking the imaginary part of the retarded self-energy, we  obtain \cite{teber}\be
\delta S(q,\omega)\approx  (K_c/12)(\alpha_-+\alpha_+)^2 q^3  \delta(\omega-v_sq),\label{spindeltapeak}
\ee
where \be \label{alphapm} \alpha_\pm=\kappa_\pm/(v_c\pm v_s).\ee 
This shows that the DCSF exhibits a narrow peak  with spectral weight of order $q^3$ at the spin mass shell $\omega=v_sq$.

The question then  is how the spin peak in Eq. (\ref{spindeltapeak}) is broadened by treating band curvature operators to higher orders. We first note that the condition $q\ll m(v_c-v_s)$ ensures that the spin peak is well separated from the charge peak. By analogy with the discussion in Sec. \ref{sec:chargepeak}, we expect that the width of the spin peak is set by  decay processes that couple degenerate states with multiple  spin bosons propagating in the same direction.  However, in contrast with the case of the charge peak,  in Eq. (\ref{H3}) there is no dimension-three spin-only  chiral operator that would be equivalent to a parabolic  dispersion about spinon Fermi points. Thus the broadening of the spin peak must come from higher-order on-shell decay processes. In fact, the leading irrelevant spin-boson interactions allowed by symmetry are  \emph{quartic} in the spin currents  \be \zeta_-(\mathbf{J}_{R/L}^2)^2 ,\zeta_+\mathbf{J}_{R}^2\mathbf{J}_{L}^2,\lambda_1 (\mathbf{J}_{R}\cdot\mathbf{J}_{L})^2, \lambda_2 \mathbf{J}_{R}\cdot\mathbf{J}_{L} (\mathbf{J}_{R}^2+\mathbf{J}_{L}^2).\label{dim4}\ee 
In principle, these operators are present as perturbations to model (\ref{eq:luttingermodel}) plus (\ref{H3}). They are also generated by ``projecting'' into a subspace with energy $|\omega-v_sq|\ll (v_c-v_s)q$ and integrating  out ``high-energy'' charge bosons. The spin part of the resulting  model for $\omega\approx v_sq$ is equivalent to the low-energy effective model for the XXZ spin chain at zero magnetic field.\cite{lukyanov} 

Perturbation theory in the dimension-four operators in Eq. (\ref{dim4}) is highly singular for $\omega\approx v_sq$.\cite{pereiraJSTAT}  Unfortunately, it is not known how to sum up the expansion in this case. Refermionization  does not solve the problem because the effective fermionic model with dimension-four operators  contains not only band curvature terms, like $\psi^\dagger\partial_x^3\psi^{\phantom\dagger}$, but also \emph{intrabranch} (i.e. which do not mix $R$ and $L$) residual interactions of the form $
\psi^\dagger\partial_x\psi^{\phantom\dagger}\partial_x\psi^\dagger\psi$, which also contribute to the broadening at leading order in $q$. Nevertheless, simple power counting tells us  that the width of the spin peak  should scale like $\delta\omega_s(q)\sim \mc{O}(q^3)$, rather than  $\mathcal{O}(q^2)$. This is consistent with the result for the DSSF of the XXZ model at zero field,\cite{pereiraJSTAT} where it is known that  the spectral weight is dominated by two-spinon excitations and the exact spinon dispersion takes the form $\epsilon_s(k)=v_s\sin(k)\approx v_s (k-k^3/6+\dots)$ about the spinon Fermi points.

\subsection{Edge singularities of the spin peak\label{sec:spinedge}} 
To be able to say more about the line shape of the spin peak, we refermionize the spin currents into \emph{interacting} spinless fermions. This is equivalent to inverting the Jordan-Wigner transformation in the continuum and writing down a SU(2) symmetric model for the fermions associated with the spin excitations. In other words, the idea is analogous to deriving the bosonic Hamiltonian for the Heisenberg spin chain by starting from the XXZ model and tuning the Luttinger parameter to the SU(2) symmetric value (a strongly interacting limit with Luttinger parameter $K=1/2$), as opposed to deriving the bosonic Hamiltonian directly from the Hubbard model (in which case the spin bosons come out noninteracting with $K_s=1$).\cite{tsvelik, Giamarchi}  The new ingredient here is that the spin excitations are coupled to gapless charge modes.

The mapping of the bare chiral fermion densities to the spin currents in Eq. (\ref{eq:luttingermodel}) is $\psi^\dagger_{s,R/L}\psi^{\phantom\dagger}_{s,R/L}=(3J^z_{R/L}-J^z_{L/R})/2$, as follows from a canonical transformation for the spin bosonic fields.  Spin inversion symmetry implies that the dispersion of these fermionic spinons is particle-hole symmetric.  We assume that the exact dispersion about the right Fermi point is given by \be \label{dispersion} \epsilon_s(k)\approx v_s k-\gamma k^3,\ee with the unknown parameter $\gamma>0$. We expect that $\gamma$  stems from dimension-four operators in the bosonic model and is of order $1/(m k_F)$.

In terms of fermions, the operator $\mathbf{J}_R^2$ that gives rise to the spin peak in Eq. (\ref{spindeltapeak}) creates particle-hole pairs on the spinon Fermi sea.  We can study the behavior near the edges of multi-spinon continua using the methods of Refs. \onlinecite{pustilnik06, pereiraPRL}. The absolute lower threshold $\omega_{s-}(q)=\epsilon_s(q)$  is defined by an excitation with a particle at the Fermi surface and a hole at momentum $-q$ below  the Fermi point. For $\omega-\omega_{s-}(q)\ll \gamma q^3$, we define a ``deep spinon''  subband by expanding\cite{pustilnik06} \be
\psi_{sR}\sim \psi_{sr}+e^{-iqx} d_s^\dagger\label{modeexp},\ee 
where $\psi^\dagger_{sr}$ creates  low-energy spinons near the right Fermi point  and $d_s^\dagger$ creates a hole at momentum $-q$ below  the Fermi point. This leads to the quantum impurity model\bea
\mc{H}_{s}^-&=&\mc{H}_{\ell}+d_s^\dagger (\omega_{s-}-iu\partial_x)d_s^{\phantom\dagger}-(V_r J_r^z+V_l J_l^z)d_s^\dagger d_s^{\phantom\dagger}\nonumber\\
&&+2\pi q(\kappa_-' J_r+\kappa_+'J_l)d_s^\dagger d_s^{\phantom\dagger},\label{xray}
\eea
where $u\approx v_s-3\gamma q^2$ is the velocity of the $d_s$ hole. The spin-only part of the quantum impurity model given by the first line in Eq. (\ref{xray}) is derived as explained in Ref. \onlinecite{pereiraPRL}, by applying the mode expansion (\ref{modeexp}) to a generic model of interacting spinless fermions with the dispersion in Eq. (\ref{dispersion}). The second line stems from the coupling of the energy density of the spinon field to the bosonized holon density.  Here $J_{r/l}$ stand for the bosonized charge currents with cutoff at energy scale $\ll \gamma q^3$, which allows the ``high-energy'' spinon to emit  low-energy charge bosons such that the energy remains near $\omega=\omega_{s-}(q)$. Note that in this procedure we keep only \emph{marginal} operators in the quantum impurity model, as irrelevant operators can only introduce subleading power-law singularities at the threshold. This is not to be confused with the presence of irrelevant operators in the original bosonic model (\ref{H3}), which are essential to argue for the nonlinearity of  holon and spinon dispersions and for the very existence of the deep spinon threshold. 

Rather than keep track of the parameters in the derivation of model (\ref{xray}), it is more useful to introduce the model phenomenologically (it contains all the marginal operators allowed by symmetry)  and to fix the coupling constants by symmetry and phenomenological relations. The parameters $V_{r/l}$   can be fixed by realizing  that the same model (\ref{xray}) can be used to calculate the lower edge singularity of the  DSSF. This is because the operator $J_{R/L}^z$  that enters the longitudinal spin-spin correlation function also creates two-spinon excitations in the fermionic representation, and the lower threshold of the support of the DSSF is also given by the deep spinon excitation. The important constraint comes from SU(2) symmetry, which imposes that the exponents for the longitudinal and transverse DSSF must coincide.\cite{imamb2} We reproduce this argument in detail in the appendix. 

The coupling constants $\kappa_\pm^\prime$ in Eq. (\ref{xray}) are related to exact phase shifts at the holon Fermi points due to the creation of a high-energy spinon. We want to show that at small $q$ these are also related to the band curvature parameters in Eq. (\ref{H3}). It is easy to show that an infinitesimal change in the chemical potential $\delta\mu$ gives rise to a shift in the energy of the high energy spinon $
\delta\omega_{s-}=\delta\mu\sqrt{K_c}(\kappa_-^\prime+\kappa_+^\prime)q/v_c$. This allows us to write\be
\kappa_-^\prime+\kappa_+^\prime=\frac{v_c}{\sqrt{K_c}q}\frac{\partial \omega_{s-}}{\partial \mu}.
\ee
But from the exact spinon dispersion we have $\omega_{s-}=v_sq+\mathcal{O}(q^3)$, hence\be
\kappa_-^\prime+\kappa_+^\prime=\frac{v_c}{\sqrt{K_c}}\frac{\partial v_{s}}{\partial \mu}+\mathcal{O}(q^2).\label{kappap1}
\ee
The second relation for $\kappa_-^\prime-\kappa_+^\prime$ can be obtained by imposing Galilean invariance to Hamiltonian (\ref{xray}). Similarly to the discussion in Section III, we compare momentum and current operators. The contribution from the spin-charge coupling terms in Eq. (\ref{xray}) to the current density (defined from the continuity equation for the charge density) is\be
\mathcal{J}_d=\sqrt{K_c}q(\kappa_-^\prime-\kappa_+^\prime)d^\dagger_sd^{\phantom\dagger}_s.
\ee 
Therefore, if we consider an excited  state in which we create a particle-hole pair of spinons with a deep hole at momentum $k_F-q$ and a particle at $k_F$, the current of this state is $\sqrt{K_c}q(\kappa_-^\prime-\kappa_+^\prime)$. Demanding that this current be equal to the momentum $q$ of the state divided by the mass $m$, we find\be
\kappa_-^\prime-\kappa_+^\prime=
1/(\sqrt{K_c}m).\label{kappap2}\ee
Comparing Eqs. (\ref{kappap1}) and (\ref{kappap2}) with (\ref{sumkappas}) and (\ref{diffkappas}), we conclude that \be
\kappa_\pm^\prime=\kappa_\pm+\mathcal{O}(q^2).\label{kappaident}
\ee

Using model (\ref{xray}) and Eq. (\ref{kappaident}), we can show (see appendix)  that the DCSF diverges at the lower edge of the two-spinon continuum as $
S(q,\omega)\sim (\omega-\omega_{s-})^{\mu_{s-}},\label{loweredgesing}$ with  exponent 
\be  \label{mu}
\mu_{s-}=-1/2+(\alpha_-^2+\alpha_+^2)q^2/2+\mc{O}(q^4),
\ee
where $\alpha_\pm$ is defined in Eq.~(\ref{alphapm}). Therefore, as $q\to0$, the exponent approaches the universal value $-1/2$, which depends only on SU(2) symmetry. The $q^2$ correction to  $\mu_{s-}$ is due to the coupling to gapless charge bosons with energy $\ll \gamma q^3$. This exponent  should be contrasted with the square-root singularity of the DSSF for the Heisenberg model. \cite{karbach} We note that $\mu_{s-}$ differs from the corresponding exponent  for SU(2) bosons at the magnon threshold, 
$\mu_{m}=-1+\mc{O}(q^2)$.\cite{zvonarev,kamenev}

The upper edge of the two-spinon continuum is given by $\omega_{s+}(q)=2\epsilon_s(q/2)$. As discussed in Ref.  \onlinecite{pereiraPRL}, near this edge the spectral weight is suppressed by resonant scattering between spinons with equal velocity. If most of the spectral weight of the spin peak is due to  two-spinon excitations, the width can be defined as \be\delta\omega_s(q)=\omega_{s+}(q)-\omega_{s-}(q)=3\gamma q^3/4.\ee  
While the upper threshold of two-spinon continuum in the integrable XXZ model exhibits a square-root cusp, here we expect that the upper threshold of the two-spinon continuum is rounded by higher order (in $q$) processes, at least for non-integrable models.

\subsection{Smearing of the charge peak: decay rate of the charge boson\label{sec:smear}}

\begin{figure}[t]
\begin{center}
\includegraphics*[width=85mm]{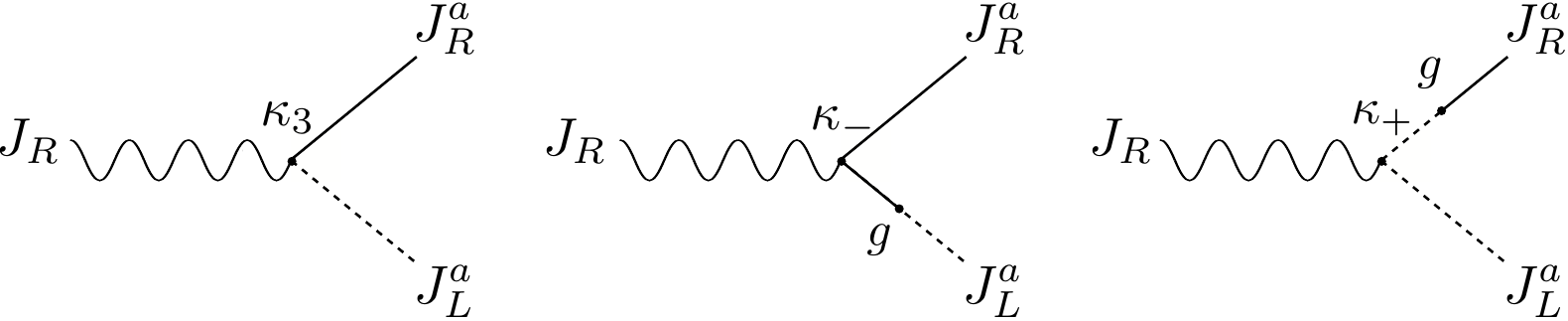}
\caption{Processes that contribute to the effective vertex $\kappa_3^{\textrm{eff}}$ in Eq. (\ref{eq:kappa3tot}) to leading order in dimension-three operators. The decay of the charge boson into a pair of right and left-moving spin bosons leads to a tail in the charge structure factor for $\omega > v_s q$. \label{fig:diagram2}}
\end{center}
\end{figure}

In general, we expect $S(q,\omega)$ to have nonzero spectral weight anywhere above the lower threshold $\omega_{s-}(q)$. A tail between the spin and charge peaks is generated due to the decay of the charge boson into a pair of $L$ and $R$ spin bosons, as depicted in Fig.~\ref{fig:diagram2}. The effective vertex is calculated from the three-point function
\bea
G(\mathbf{k} _1,\mathbf{k} _2,\mathbf{k} _3)&\equiv&\langle (J_L+J_R)(\mathbf{k} _1)J_L^a(\mathbf{k} _2)J_R^b(\mathbf{k} _3) \rangle  \nonumber \\
&=&\frac{4\pi^2}{3} \kappa_{3}^{\textrm{eff}} \delta^{a b} (C_L+C_R)(\mathbf{k} _1) S_L(\mathbf{k} _2)  \nonumber\\
&&\times  S_R(\mathbf{k} _3)(2 \pi)^2 \delta(\mathbf{k} _1+\mathbf{k} _2+\mathbf{k} _3),
\eea
where $\mathbf{k}=(k,\omega)$ is a two-momentum. To first order in $\kappa_\pm,\kappa_3$ (leading order in $q/k_F$ in the contribution to $S(q,\omega)$), there are two contributions to the effective vertex, one from  $\kappa_3$ and the other from a combination of $\kappa_\pm$ and $g$. 
We find
\begin{equation}
\label{eq:kappa3tot}
\kappa_{3 }^{\textrm{eff}}=\kappa_3 +\frac32 g (\kappa_- + \kappa_+).
\end{equation}

Away from spin and charge mass shells, i.e. for $|\omega-v_sq|\gg \delta\omega_s(q)$ and $|\omega-v_cq|\gg \delta\omega_c(q)$,  the tail of the spin peak can be calculated by second order perturbation theory in $\kappa_3^{\textrm{eff}}$. Similarly to the calculation in Sec.~\ref{se:spinpeak}, we obtain a correction to the charge boson propagators
\bea
\label{Ckappa3}
\delta C^{\kappa_3^{\textrm{eff}}}_{\alpha \beta}(q,i \omega)&=& -\frac{16\pi^4}3 (\kappa_{3}^{\textrm{eff}})^2 C_\alpha(q,i \omega) \nonumber\\
&&\times C_\beta(q,i \omega) \Pi_{RL}(q,i \omega),
\eea
where $\alpha,\beta = R,L$ and $\Pi_{RL}(q,i \omega)$ is the self-energy with one right-moving and one left-moving spin boson, as defined in Eq.~(\ref{eq:Pi}). The calculation of $\Pi_{RL}$ yields\cite{pereiraJSTAT}
\bea
\label{Ckappa3A}
\Pi_{RL}^{\textrm{ret}} (q,\omega)=\frac{1}{32 \pi^3} \left[\frac{\Lambda^2}{2 v_s} - \frac{\omega^2 - v_s^2 q^2}{8 v_s^3}\right. \nonumber \\
\times\left. \log  \frac{(v_s q)^2 - (\omega+ i \eta)^2}{4 v_s^2 \Lambda^2} \right].
\eea
While the real part is ultraviolet divergent, the imaginary part is not. The imaginary part gives the tail  in the DCSF
\be
\delta S(q,\omega)\approx \frac{K_c (\kappa^{\textrm{eff}}_{3})^2 }{24  v_s^3} \left[\frac{v_cq^2}{\omega^2-v_c^2q^2}\right]^2(\omega^2-v_s^2q^2),\label{spintail}
\ee 
for $\omega>v_sq$ and $|\omega-v_c q|\gg \eta_-q^2$. 

At this point, we recall that both $g$ and $\kappa_3$ that appear in the amplitude for $\kappa_3^{\textrm{eff}}$ scale logarithmically with the infrared cutoff (see Sec. \ref{se:RG}). Within RG improved perturbation theory, the bare  $g$ and $\kappa_3$ in Eq. (\ref{eq:kappa3tot}) are replaced by the renormalized ones in Eqs. (\ref{gscaling}) and (\ref{zetaln2}), with cutoff set by the small momentum $\Lambda\sim q$. In the long-wavelength limit, such that $g(q)\sim 1/\ln(k_F/q)\ll 1$, we have\bea
g(q)&\sim& 1/\ln(k_F/q)\\ \kappa_3(q)&\sim &(\kappa_3/g^2) /\ln^2(k_F/q).
\eea
At leading logarithmic order, we can drop the contribution from $\kappa_3$ in $\kappa^{\textrm{eff}}_3(q)$ and Eq. (\ref{spintail}) becomes\be
\delta S(q,\omega)\approx \frac{3K_c [g(q)]^2 }{32  v_s^3} \left[\frac{v_c(\kappa_-+\kappa_+)q^2}{\omega^2-v_c^2q^2}\right]^2(\omega^2-v_s^2q^2).\label{taileq}
\ee

The DCSF was calculated by similar methods in Ref. \onlinecite{teber}, but the tail between the spin and charge peaks was not obtained because backscattering processes ($g$ and $\kappa_3$ in our notation) were neglected.

The presence of the tail means that the charge peak discussed in Sec. \ref{sec:chargepeak} is inside a continuum of spin excitations. The coupling to the continuum results in a decay rate for the charge excitations. The imaginary part of the self-energy $\Pi_{RL}$ can be absorbed into the charge boson propagator in the form \be
C^{\textrm{ret}}_{R/L}(q,\omega) = \frac{1}{4 \pi}\frac{\pm q}{\omega\mp v_c q + i  /\tau_c},\ee
where $\tau_c^{-1}(q)$ is the decay rate given by
\be
\label{tauc}
\frac1{\tau_c}=\frac{\pi  (\kappa^{\textrm{eff}}_{3})^2 (v_c^2-v_s^2)q^3}{ 192  v_s^3}.
\ee
In the limit $g(q)\sim 1/\ln(k_F/q)\ll1$, we obtain\be
\frac1{\tau_c}=\frac{3[g(q)]^2\pi  (\kappa_-+\kappa_+)^2 (v_c^2-v_s^2)q^3}{ 256  v_s^3}.\label{tauc2}
\ee
For $q\ll m(v_c-v_s)$, the decay rate in Eq. (\ref{tauc2}) is smaller than $\delta\omega_c(q)\sim q^2$, which is due to decay of the charge boson within the charge sector (see Sec. \ref{sec:chargepeak}).  However, $\tau_c^{-1}$ is important because it  is responsible for rounding off the edges of the charge peak. This can be confirmed by calculating the decay rate for a single high-energy  holon -- a $d_c$ particle in the quantum impurity model for the edges of the two-holon continuum, similar to the calculation in Ref. \onlinecite{pereiraPRB} for the spinless case.. The decay rate is  due to the perturbation \be \label{effectivetheory}
\delta\mathcal{H}_3\sim \xi_3d_c^\dagger d_c^{\phantom\dagger} \partial_x\varphi^s_R\partial_x\varphi_L^s.
\ee
The parameter $\xi_3$ gives the amplitude for a process in which a holon scatters off two spinons moving in opposite direction. This is \emph{not} a three-electron scattering process and in principle $\xi_3\neq 0$ even for integrable models. Using the result in Eq. (5.8) of Ref.  \onlinecite{pereiraPRB}, we obtain\be
\frac{1}{\tau_c}\propto \frac{(\xi_3)^2(v_c^2-v_s^2)[\Lambda(q)]^3}{v_c^3},
\ee
where $\Lambda(q)$ is the cutoff of the high-energy subband. Setting $\Lambda(q)\sim q$, we recover the momentum and interaction dependence of the decay rate in Eq. (\ref{tauc2}) if we assume that $\xi_3$ does not vanish as a power law of $q$ in the limit $q\to 0$. From comparison with Eq. (\ref{tauc}), we expect $\xi_3\propto \kappa_3^{\textrm{eff}}$. This should be contrasted with the spinless case, where the coupling constant in Eq. (5.13) of Ref. [\onlinecite{pereiraPRB}] has to vanish like $q^2$ because of statistics, since there is no $s$-wave scattering for spinless fermions. For the spinful case,  statistics alone does not imply that the amplitude $\xi_3$ vanishes as $q^2$ or high powers of $q$.   

Note also that the result for $1/\tau_c$ in Eq. (\ref{tauc2}) is nonperturbative in the electron-electron  interaction, since for $q\ll m\tilde{V}_0\ll k_F$, we have $\tau_c^{-1}\sim g^2\tilde{V}_0q^3/k_F^2$, which is \emph{third} order in the interaction strength. Furthermore, this result implies that,  
even  for an integrable model, the power-law singularities\cite{pustilnik06}   at $\omega_{c\pm}$ are removed at order $q^3$. The same decay rate $1/\tau_c$ rounds off the singularity at the holon mass shell in the electron spectral function.\cite{voit}  This is remarkably different from the spinless case, where it is believed that integrable models can have exact singularities above the lower threshold because the decay rate of a high-energy particle may vanish exactly.\cite{pereiraPRB}

We note that if   the phenomenological relations Eqs.~(\ref{sumkappas}) and (\ref{kappa3}) are valid  for the running coupling constants and we substitute them in Eq. (\ref{eq:kappa3tot}), we find $\kappa_3^{\textrm{eff}}=-(3/2)v_cv_sK_c^{-1/2}\partial g/\partial\mu$.\cite{glazmanetal} At low energies, $g(\Lambda)\sim g(\Lambda_0)/[1+g(\Lambda_0)\ln (\Lambda_0/\Lambda)]$. It is not clear how imposing the phenomenological relations at all energy scales can be reconciled with the result from the RG. It is important for our results in Eqs. (\ref{taileq}) and (\ref{tauc2}) that even if  $\partial g/\partial\mu=0$ at some energy scale, the effective vertex $\kappa_3^{\textrm{eff}}$ will be generated by the RG flow because $\kappa_3$ and $g$ scale differently, and the leading logarithmic dependence is due to the $g$ term  in Eq. (\ref{eq:kappa3tot}). 

\begin{figure}[t]
\includegraphics*[width=.9\hsize,scale=0.95]{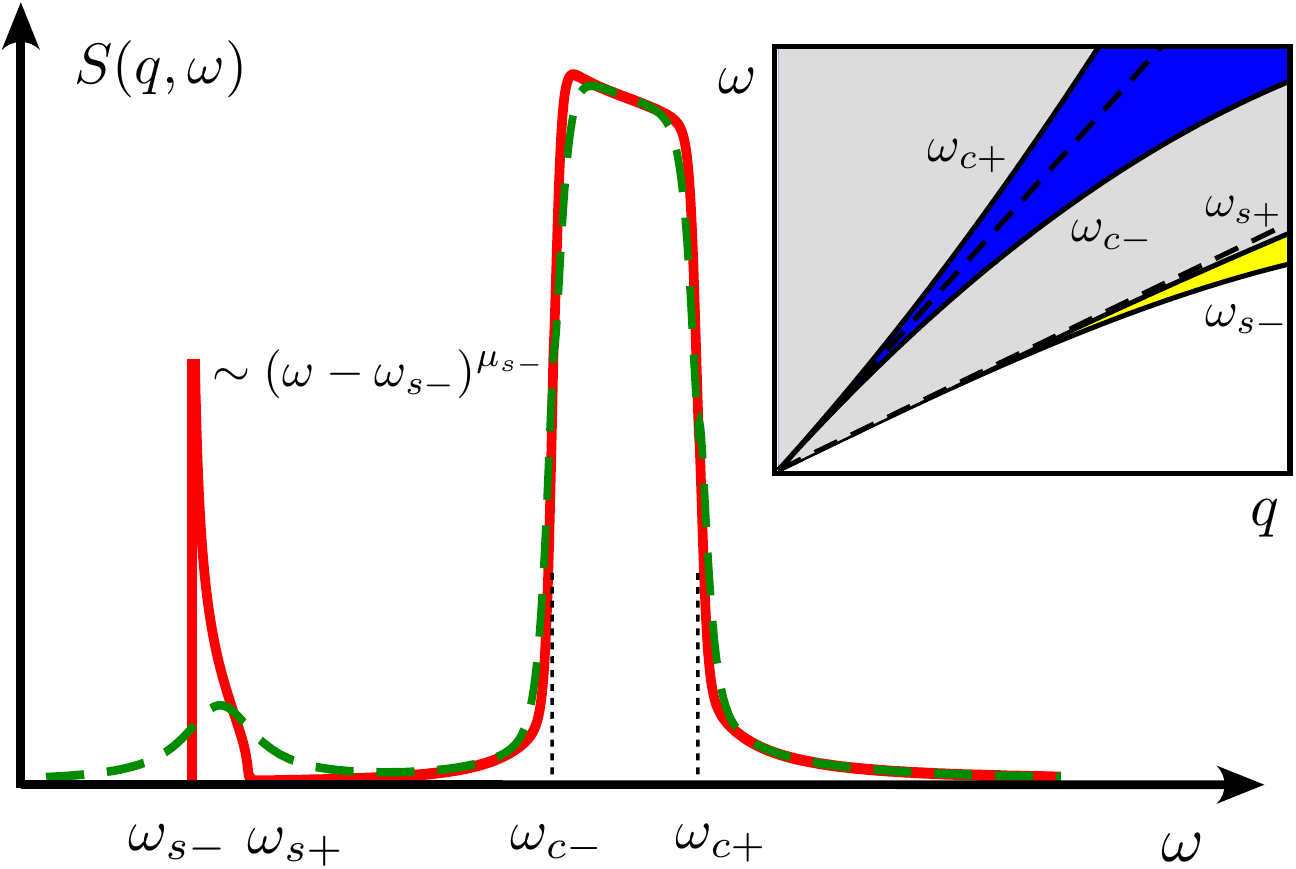}
\caption{(Color online) Schematic  excitation spectrum (inset on the upper right corner) and line shape (main figure) for the DCSF of spin-1/2 fermions at small $q$. The dashed lines in the $(q,\omega)$ plane represent the linear dispersion of charge and spin modes in the Luttinger model. The two-holon and two-spinon continua are   represented by blue and yellow regions bounded by $\omega_{c\pm}(q)$ and $\omega_{s\pm}(q)$, respectively. At zero temperature and in the regime $q\ll m(v_c-v_s)$, the line shape of $S(q,\omega)$ (solid red line in the main figure)  has well defined peaks inside the two-holon and two-spinon continua. At finite temperature $\delta\omega_s(q)\ll g^2T\ll v_cq$, diffusion broadens the spin peak  into a Lorentzian (dashed green line).\label{fig:lineshape}}
\end{figure}

Finally, putting together all the pieces, we construct the final picture for the DCSF at zero temperature in Fig. \ref{fig:lineshape}. Note, in particular, that there is only a rounded threshold  at $\omega\approx \omega_{c-}$. 

\subsection{Finite temperature effects\label{sec:finiteT}}

We now discuss the effects of finite temperature on the broadening of the charge  and spin peaks in the DCSF. This will be important to compute the temperature dependence of the drag resistivity in Eq. (\ref{dragresist}). 

We consider the regime where both temperature and band curvature energy scale are small compared to the scale of spin charge separation: $T \ll (v_c-v_s)v_c/\eta_-$ and $q \ll (v_c-v_s)/\eta_-$. Assuming that $v_c$, $v_s$, and $v_c-v_s$ are all of the order of $v_F$ and $1/\eta_-$ is of the order of $m$, these conditions mean roughly  $T \ll T_F = m v_F^2/2$ and $q \ll k_F$.
This is the regime in which we may expect the spin and charge peaks to remain well separated. Neglecting the overlap between the spin and charge peaks, the line shape of the charge peak  can be approximated by the finite temperature result for the imaginary part of the density-density correlation function  for  free fermions with mass $\sqrt{2}/\eta_-$ \cite{pustilnik,aristov} \be
A(q,\omega,T)\approx \frac{\sqrt2 K_c}{\eta_-q}[n_F(w_+)-n_F(w_-)],\label{Acharge}
\ee
where $n_F(\omega)=1/(1+e^{\omega/T})$ is the Fermi-Dirac distribution function and $w_\pm\equiv[\left(\omega\pm\delta\omega_c/2\right)^2-(v_cq)^2]/(2\delta\omega_c)$, with $\delta\omega_c=\delta\omega_c(q)=\eta_-q^2/\sqrt2$.
The width of the charge peak at finite temperature is then of the order of $\textrm{max}\{\eta_-q^2,\eta_-qT/v_c\}$. 

The calculation of the width of the spin peak is more complicated because we do not have an approximation in terms of noninteracting spinless fermions. For the purpose of calculating the drag resistivity in Eq. (\ref{dragresist}), we are only interested in whether for fixed small $q$ and at low temperature the spin peak can become broader than the charge peak. On the one hand, the broadening due solely to band curvature must be of order $(\gamma q^2/v_s)T$. As long as $q,T/v_s\ll m(v_c-v_s),\eta_-/\gamma$, this is  small compared to the broadening of charge peak. On the other hand, thermal effects have a stronger effect on spin excitations because the latter are damped by diffusion.\cite{balents,polini}

\begin{figure}[t]
\includegraphics*[width=.85\hsize,scale=0.95]{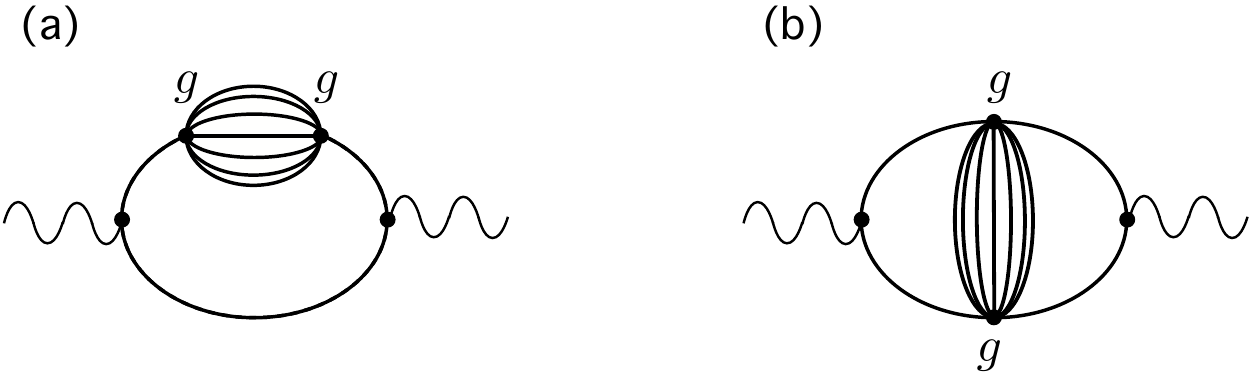}

\caption{Diagrams at $\mc O(\kappa_\pm^2g^2)$ in the self-energy for the charge boson: (a) self-energy correction to the spin boson propagator; (b) vertex correction. The bubble with multiple lines denotes the correlation function for the operator $J_R^+J_L^-\sim e^{i\sqrt{4\pi}(\varphi_R^s-\varphi_L^s)}$, following Ref. \onlinecite{oshikawa}. \label{fig:damp}}
\end{figure}

Recall that the spin peak stems from the self-energy with two spin bosons propagating in the same direction (see Fig. \ref{fg:diagram1}).
We can calculate the finite temperature broadening by neglecting band curvature operators and applying perturbation theory in the marginally irrelevant operator $g$ in Eq. (\ref{eq:luttingermodel}). As we did in section \ref{sec:smear}, we neglect the $\kappa_3$ vertex in the leading logarithmic approximation. To order $(\kappa_\pm g)^2$, there are two types of diagrams in the self-energy for the charge boson, as illustrated in Fig. \ref{fig:damp}. The first type amounts to a self-energy correction to the spin boson propagator. The transverse part of the perturbation, $-\pi v_s g(J_L^+J_R^-+h.c.)$ gives rise to a nonzero imaginary part of the retarded self-energy, which can be calculated following Ref. \onlinecite{oshikawa}.
We can sum up the series for this type of diagram by defining the dressed spin propagator\be
\tilde S_{R}(q,i\omega)=\frac{1}{4\pi}\frac{q}{i\omega-v_sq-\Sigma(q,i\omega,T)}.
\ee
The other type of diagram (Fig. \ref{fig:damp}(b)) is a vertex correction. Since the thermal broadening of the spin peak is already obtained within the approximation of keeping only self-energy-type diagrams like the one in Fig.  \ref{fig:damp}(a), we will make the approximation of neglecting vertex corrections. By doing this, the two-spin-boson correlation function becomes \bea
(4 \pi)^2 \Pi_{RR}(q,i\omega)&=&-\int _{-\infty}^\infty \frac{dq^\prime}{2\pi}\,T\sum_{i\nu_n}\tilde S_R(q^\prime,i\nu_n)\nonumber\\
&&\times\tilde S_R(q-q^\prime,i\omega-i\nu_n)\nonumber\\
&\approx&\int _{-\infty}^\infty \frac{dq^\prime}{2\pi}\frac{q^\prime(q-q^\prime)}{i\omega-v_sq-\Sigma_{q^\prime}-\Sigma_{q-q^\prime}}\nonumber\\
&&\times[n_B(v_sq^\prime)-n_B(v_sq^\prime-v_sq)],\label{sigmaapprox}
\eea
where $n_B(\omega)=1/(e^{\omega/T}-1)$ and $\Sigma_q=\Sigma_q(T)=\Sigma(q,v_sq,T)$.
The decay rate for the spin boson at finite temperature is the well-known spin current relaxation rate\cite{balents} \be
\frac1{\tau_s(T)}=-\textrm{Im }\Sigma^{\textrm{ret}}_q\approx \frac{\pi }2[g(T)]^2T,
\ee
where  $g(T)\approx g/[1+g\ln(T_F/T)]$, with $T_F\sim mv_F^2$, is the  coupling constant at scale $T$. The finite temperature result for the spin peak in this approximation is then  \be
\delta A(q,\omega,T)\approx   \frac{K_c}{48\pi}\frac{(\alpha_-+\alpha_+)^2q^3F(q,T)\tau_s}{1+[(\omega-v_s q)\tau_s/2]^2},\label{deltaAspin}
\ee
where \bea
F(q,T)&=&6\int_{-\infty}^\infty du\,u(1-u)\nonumber\\
&&\times[n_B(v_squ)-n_B(v_sq(u-1))],
\eea
such that $F(q,T\to0)=1$.
Since we neglected band curvature effects in the spin boson propagator, this approximation is only valid for $1/\tau_s(T)\gg \gamma q^3$.  In Eq. (\ref{sigmaapprox}), we also assumed $1/\tau_s(T)\ll v_sq$. Therefore we expect that the line shape of the spin peak at finite temperature be well described by a Lorentzian for a range of $q$ that scales linearly with temperature, $q\sim T/v_s$, such that $T/g(T)\ll (v_s^3/\gamma)^{1/2}$. 

More generally, for fixed $q$ there is a crossover temperature $T^*$ given by the condition $1/\tau_s(T^*)\sim \gamma q^3$ at which the line shape of the spin peak goes from highly  asymmetric (with a peak near the zero temperature threshold) below $T^*$ to approximately Lorentzian in the diffusion-dominated regime above $T^*$.

\section{Application to the drag resistivity}

Finally, as an application of our results for the DCSF, in this section we will discuss the density and temperature dependence of the drag resistivity in Eq. (\ref{dragresist}). Here we restore the index $i=1,2$ for properties of the drive wire and drag wire, respectively. Note that Eq. (\ref{dragresist}) is symmetric under the exchange of drive and drag wires. We may then assume $v_{c1}\leq v_{c2}$. We envision an experiment in which the electron density of the drag wire is varied while the one in the drive wire is kept fixed. In principle, this could be achieved with vertically coupled wires whose densities are tuned independently by top and back gates.
We can think that the sharply peaked DCSF of the drive wire is the reference for integrating Eq. (\ref{dragresist}) in the $(q,\omega) $ plane, and approximate\bea
r&\approx&\frac{U^2}{4\pi^3 \nu_1 \nu_2T}\int_0^\infty dq\,\frac{q^2}{\sinh^2(v_{c1}q/2T)} \nonumber\\
&&\times\int_0^\infty d\omega\,A_1(q,\omega)A_2(q,\omega).\label{dragresist2}
\eea

Clearly, the forward scattering contribution to the drag resistivity is maximum for ideal density matching, $v_{c1}=v_{c2}$,\cite{comment} in which case 
it is dominated  by the overlap of the charge peaks of the two wires over modes with $\omega\approx v_{ci}q\sim T\ll T_F$. The temperature dependence  in this case is the same as for spinless fermions,\cite{pustilnik} $r\sim T^2$  for $m|v^2_{c1}-v_{c2}^2| \ll T \ll T_F$. Also like the result for spinless fermions, a much weaker response, $r\sim T^5$, is obtained for general density mismatch in the regime  $T\ll m|v^2_{c1}-v_{c2}^2|$, from the overlap of the charge peak for one wire with the  tails of the DCSF for the other wire. 

The new effect due to the spin degree of freedom is related to the presence of a spin peak in the DCSF illustrated in Fig. \ref{fig:lineshape}. It suggests that the drag resistivity can be enhanced  when the electron densities in the wires are rather different  but the charge peak of one wire overlaps with the spin peak of the other wire. This happens over an extended region in the $(q,\omega)$ plane if $v_{c1}\approx v_{s2}$. The interpretation is that under this condition holons in wire 1 can efficiently scatter off spinons in wire 2, which in their turn transfer momentum to holons in the same wire.

Let us discuss the temperature dependence of the drag resistivity when $v_{c1}=v_{s2}$.  We are interested in the regime $T\ll mv_{c1}(v_{c2}-v_{c1})$, where   spin charge separation may be measurable. According to the result in section \ref{sec:finiteT},  the spin peak $\delta A(q,\omega,T)$ for values of $q$ that scale linearly with $T$ assume a Lorentzian line shape at low temperatures such that  $T/g_2(T)\ll (v_{s2}^3/\gamma_2)^{1/2}$. The width of the charge peak for wire 1 for $q\sim T/v_{s2}$ is of order $\eta_{-,1}T^2/v_{s2}v_{c1}$. If the temperature is also low enough that $T/[g_2(T)]^2\ll v_{c1}v_{s2}/\eta_{-,1}\sim T_F$, the spin peak for wire 2 is broader than the charge peak for wire 1. Using Eq. (\ref{Acharge}) for wire 1 and Eq. (\ref{deltaAspin}) for wire 2, we find that in this regime the drag resistivity in Eq. (\ref{dragresist2}) scales like $r\sim T^5/[g_2(T)]^2$. However, if the value of $g_2(T)$ is  small even at intermediate temperatures, there will be in general a temperature regime $ g_2(v_{s2}^3/\gamma_2)^{1/2}\ll T\ll mv_{c1}(v_{c2}-v_{c1})$ where diffusion is not effective, in the sense that for $q\sim T/v_{s2}$ the broadening of the spin peak due to diffusion is smaller than the one due to band curvature. In this case, the spin peak for wire 2 is  narrower than the charge peak for wire 1 for the same value of $q\ll k_{Fi}$. Therefore, the integral in Eq. (\ref{dragresist2}) can be evaluated by considering that the entire spectral weight of the spin peak is inside the charge peak. In this regime,  the drag resistivity in Eq. (\ref{dragresist2}) scales like $r\sim T^4$. In summary,\be
r\sim \left\{\begin{array}{lc}T^5\ln^2\frac{T_F}T,& T\ll g_2(T) \sqrt{\frac{(v_{s2})^3}{\gamma_2}},[g_2(T)]^2T_F\\
T^4,& {g_2} \sqrt{\frac{(v_{s2})^3}{\gamma_2}}\ll T\ll mv_{c1}(v_{c2}-v_{c1}).
\end{array}\right.
\ee
Notice that in the low temperature limit diffusion suppresses the drag by making the spin peak broader than the charge peak. Nevertheless, even in the spin diffusion regime  the smallness of $g_2(T)\sim 1/\ln(T_F/T)$ makes the drag  larger than the background  contribution from the tails of the DCSF, which scales like $r\sim T^5$, as discussed above.

In order to compute the expression in Eq. (\ref{dragresist2}), we have used the same strength of the electron-electron interaction $\tilde V_{0}$ and $\tilde V_{2k_F}$ for both  wires. We estimated the parameters of the DCSF in each wire using the phenomenological relations in section III, expanding to first order in the interaction.  The finite temperature $A_i(q,\omega)$ are approximated by the sum of a dominant charge peak given by Eq. (\ref{Acharge}) and a smaller spin peak given by the Lorentzian in Eq. (\ref{deltaAspin}).  For modes with $q\sim T/v_{c1}$ that contribute to the integral in Eq. (\ref{dragresist2}) but are not in the regime $\gamma q^3\ll 1/\tau_s(T)$, the Lorentzian is not a good approximation for the line shape of the spin peak (which must become asymmetric with a peak near the lower edge of the two spinon spectrum). However, the drag resistivity is not sensitive to the detailed line shape since in this regime the spin peak is much narrower than the charge peak for the same $q$ and what matters most is whether the spin peak for the drag wire  falls inside the charge peak for the drive wire. 

\begin{figure}[t]
\includegraphics*[width=.95\hsize,scale=0.95]{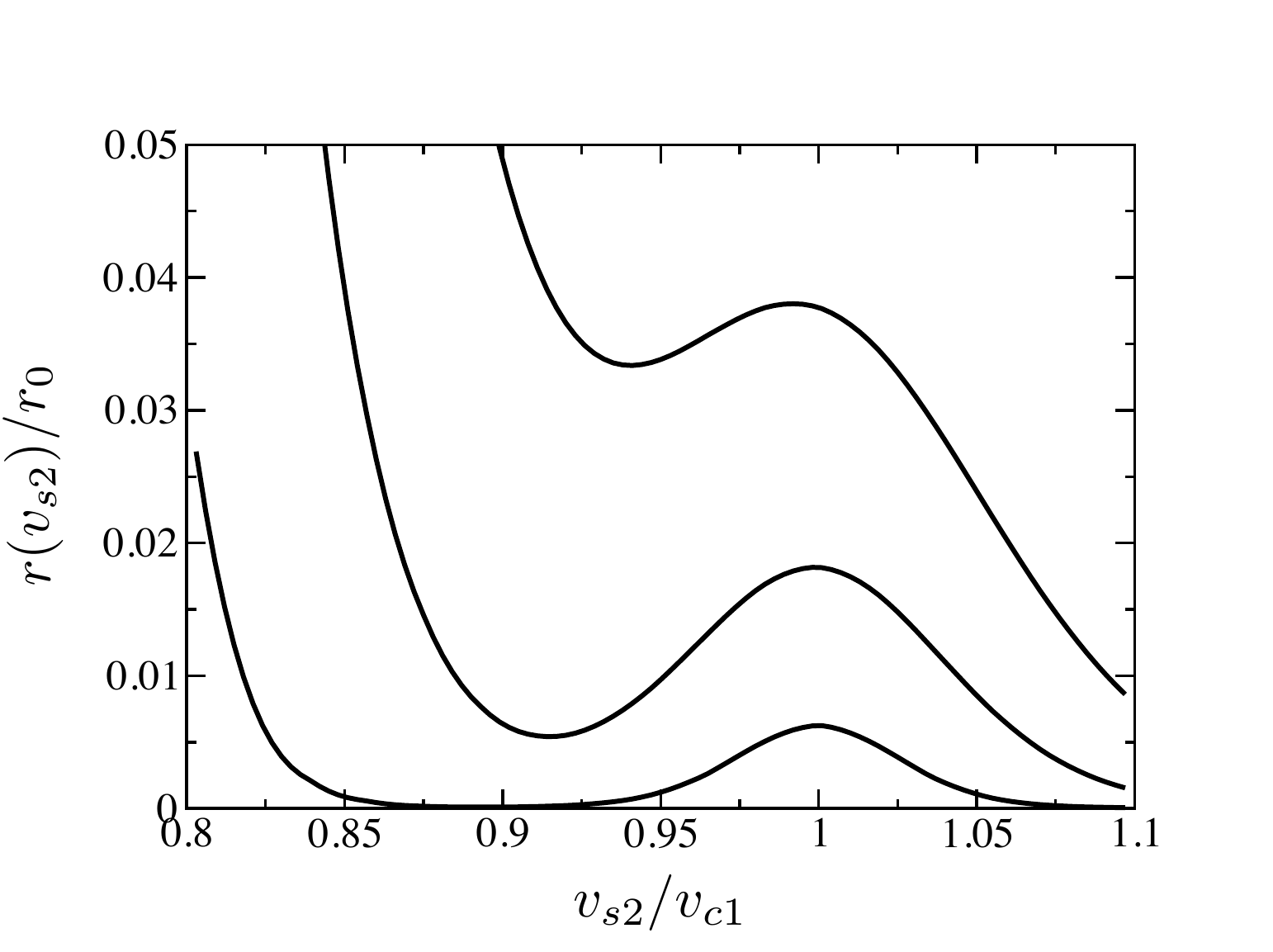}

\caption{
Drag resistivity as a function of the spin velocity $v_{s2}$ in the drag wire for three values of temperature: $T/(k_{F1}^2/m)=0.03,0.05,0.07$ (bottom to top). In this graph we set $\tilde{V}_0/(\pi v_{F1})=0.4$ and $\tilde{V}_{2k_F}/(\pi v_{F1})=0.1$  in the weak coupling expressions for $v_{ci}$, $v_{si}$, $K_{ci}$, $\eta_{-i}$, $\kappa_{\pm i}$ for wires $i=1,2$, using equations in section III.  The value of $r(v_{s2})$ is normalized by the value at zero density mismatch, $r_0=r(v_{s2}=v_{s1})$, for each temperature. Notice the peak in $r$ when the spin velocity of the drag wire matches the charge velocity of the drive wire, $v_{s2}\approx v_{c1}$.  \label{fig:drag}}
\end{figure}

Fig. \ref{fig:drag} illustrates the dependence of the drag resistivity on the density mismatch between the two wires, parametrized by the ratio $v_{s2}/v_{c1}$. The effect of spinon-assisted Coulomb drag is observed as a small peak in the drag resistivity    when the wire densities are such that $v_{s2}\approx v_{c1}$.  The height of the peak relative to the dominant response at zero density mismatch ($v_{s2}\approx v_{s1}$) increases with increasing temperature. This is because, as temperature increases, modes with larger $q$, for which the spin peak in the DCSF have relatively larger spectral weight, start to contribute to the integral in Eq. (\ref{dragresist2}). On the other hand, increasing temperature also broadens the  dominant peak  observed at $v_{s2}\approx v_{s1}$, which eventually  obscures the smaller peak at $v_{s2}\approx v_{c1}$. Therefore, it seems that the contribution of spinons to Coulomb drag would most likely be observed as a shoulder in the density dependence of the drag resistivity at intermediate temperatures.

\section{Conclusions}

We have studied the dynamic charge response for spin-1/2 fermions in one dimension beyond the usual linear dispersion approximation of Luttinger liquid theory. Unlike the spinless case,  the limits of small momentum $q$ and weak electron-electron interactions $\tilde{V}_0$ do not commute. This is due to the interplay of spin charge separation and band curvature effects. The problem of calculating the dynamical charge structure factor  for electrons with mass $m$ in the regime $q\ll m\tilde V_0$ cannot be approached by perturbation theory in the electron-electron interactions. We have used a bosonized Hamiltonian and discussed the effects of irrelevant perturbations associated with band curvature. We determined phenomenological relations for the coupling constants of these perturbations, including the ones that couple charge and spin dynamics. The renormalization group equations for the irrelevant operator denoted by $\kappa_3$, which couples charge and spin and mixes right and left-moving spin modes, shows that its effective coupling constant has a nontrivial logarithmic scaling in the low energy limit.  

Based on a picture in which collective charge and spin modes can be refermionized into spinless fermions (holons and spinons, respectively) with nonlinear dispersion, we presented an approximate line shape for the dynamic charge structure factor valid in the long wavelength limit. As a function of frequency, the dynamic charge structure factor has a dominant charge peak associated with two-holon excitations, whose width scales like $q^2$. However, the spectral weight extends down to a lower threshold described as a two-spinon excitation. We calculated the exponent of the power-law singularity at this lower threshold and found that in the limit $q\to 0$ it converges to the universal value $\mu_{s-}=-1/2+\mc{O}(q^2)$, which depends only on spin SU(2) symmetry. We expect that the spectral weight near this lower threshold is largest within a two-spinon continuum, giving rise to a spin peak in the charge structure factor. There is also a tail of the spin peak above the two-spinon continuum. The coupling between spin  and charge in the vicinity of the charge mass shell gives rise to a decay rate for the holon at order $q^3$. This decay rate is responsible for rounding off singularities at intermediate thresholds such as the edges of the two-holon continuum, regardless of the integrability of the model. 

At finite temperature, an important difference between charge and spin excitations is that the latter are damped by diffusion. In the dynamic charge structure factor this effect is manifest in the Lorentzian broadening of the spin peak in the regime where the spin current decay rate is large compared to the band curvature energy scale for the spinons. 

These results allowed us to calculate the Coulomb drag response between two quantum wires, taking into account the spin degree of freedom. In comparison with the result for spinless fermions studied in Ref. \onlinecite{pustilnik},  there is an additional effect (spinon-assisted Coulomb drag) due to spin-charge coupling: at low temperatures the drag resistivity as a function of density mismatch has a peak when the charge velocity of one wire matches the spin velocity of the other. The temperature dependence of this drag peak  has signatures of spin diffusion. 

Note: After this work had been submitted, Ref. \onlinecite{schmidt} appeared with related results for the exponent of edge singularities at arbitrary momenta.

\acknowledgments

We acknowledge helpful discussions with J. M. P. Carmelo,  L. I. Glazman, A. Imambekov and T. Schmidt.
This research was  supported by the NSF under Grant No. PHY05-51164 (R.G.P.), by NSERC (E.S.) and the A. v. Humboldt Foundation (E.S.).

\appendix
\section{Calculation of the lower edge exponent}
We provide details for the calculation of the exponent of the power law singularity in the DCSF at $\omega=\omega_{s-}(q)$ at zero temperature. While the asymptotic value $\mu_{s-}(q\to0)=-1/2$ is universal and depends only on SU(2) symmetry, following the argument of Ref. \onlinecite{imamb2}, we also obtain the $q^2$ correction due to spin-charge coupling.

As mentioned in Sec. \ref{sec:spinedge}, the lower edge of the spectrum is the same for the DCSF and for the DSSFs,  defined as \be
S^{ab}(q,\omega ) = \int_0^L dx\, e^{-iqx}\int_{-\infty }^{+\infty } dt\,  e^{i \omega t}\bra S^a(x,t) S^b(0,0) \ket. \ee 
Here $\mathbf{S}(x) =\Psi^\dagger(x) \frac{{\bm\tau}}{2} \Psi(x)$, with ${\bm\tau}$ the vector of Pauli matrices, is the spin density operator. Its long wavelength components are represented by $\mathbf{S}=\mathbf{J}_L + \mathbf{J}_R$.  For fixed $q$, the lower threshold below which the DSSFs (both longitudinal and transverse) vanish is controlled by a deep spinon excitation with energy  $\omega_{s-}(q)=\epsilon_s(q)$. 

The components of $\mathbf{S}$ satisfy SU(2) commutation relation \be [S^a(x),S^b(x')]=i \varepsilon^{abc} S^c(x) \delta(x-x'). \ee We will refermionize the spin density to spinless fermions. This can be done using an inverse Jordan-Wigner  transformation in the continuum
\bea S^z(x)  &=& \Psi_{s}^\dagger(x) \Psi_{s}(x) +\textrm{const.}, \label{JW1}\\
 S^+(x) &=&   \Psi^\dagger_{s}(x) \,e^{i \pi \int_{-\infty}^\infty dx^\prime\,\theta(x-x^\prime)S^z(x^\prime)},\label{JW2}\eea
where $\Psi_s$ is a spinless fermionic spinon field, which obeys anticommutation relations $\{\Psi^{\phantom\dagger}_s(x),\Psi^\dagger_s(x^\prime)\}=\delta(x-x^\prime)$, and $\theta(x)$ is the left-continuous Heaviside step function with $\theta(0)=0$. 

A generic spin Hamiltonian that is a function of the local spin density and respects SU(2) symmetry takes the form\be
\mathcal H_s=C_1\mathbf{S}\cdot \mathbf{S}+C_2\partial_x\mathbf{S}\cdot\partial_x \mathbf{S}+\dots,
\ee
where $\dots$ stands for higher-order irrelevant operators. By means of Eqs. (\ref{JW1}) and (\ref{JW2}), this maps onto a model of interacting spinons. The precise form of the spinon Hamiltonian is not essential here, but it must be such that in the low energy limit it yields the same equal-time correlation functions as the spin part of the Luttinger model (\ref{eq:luttingermodel}). This is directly accomplished if model (\ref{eq:luttingermodel}) is recovered by bosonization of $\Psi_s$, analogously to the bosonization of the XXZ model.\cite{Giamarchi} In this approach, the SU(2) symmetric model corresponds to strong interactions between spinons. Here we assume that the effective model for the spinons in the metallic case can be approached in the same way as the effective model for the Heisenberg spin chain, namely by starting from a generalized model of weakly interacting spinless fermions where we can expand the dispersion about the Fermi points to bosonize the low-energy degrees of freedom. SU(2) symmetry is only imposed at the end, on the results for the spin-spin correlation functions, to fix the parameters of the effective model. (It is conceivable that the strength of the spinon-spinon interaction could be tuned in a microscopic model for a metal with spin U(1) symmetry and that such weakly interacting limit could be realized.) In addition to the interactions in the effective spinon model, we must account for the coupling to gapless charge modes. For the purpose of deriving the exponent at the spinon edge, the latter can be described by bosonic fields $\varphi^c_{R/L}$ at all steps and need not be refermionized. The spin-charge coupling is then equivalent to spin-phonon coupling in spin chains. 

Let us consider that in the ground state the spinons form a Fermi sea with a particle-hole symmetric band (due to spin inversion symmetry $S^z\to-S^z$). The elementary $S^z=0$ excitations are particle-hole pairs in the spinon Fermi sea. Similarly, there are particle-hole excitations in the holon Fermi sea, but in our low-energy effective model for the lower edge  these are treated as charge bosons (since we can neglect band curvature for the holons). This picture is supported by the Bethe ansatz solution of the Yang-Gaudin or Hubbard models. We assume that we can start from a model of noninteracting spinons with dispersion $\epsilon_{s,R/L}^{(0)}(k)\approx \pm (v_s^{(0)}k-\gamma^{(0)}k^3+\dots)$ about the Fermi points $\pm k_F$. That the spinon Fermi wave vector is given by $k_{Fs}=k_F$ follows from the periodicity of the spin excitation spectrum, which is gapless at momentum $2k_F$.\cite{kamenev}
Another interpretation is that in this approach the number of spinons,\be
\int_{0}^{L}dx\, \Psi_s^\dagger(x)\Psi_s^{\phantom\dagger}(x)=k_{Fs}L/\pi =N/2,
\ee 
is fixed by the condition that the state constructed by adding (removing) $N/2$ spinons to the ground state is a fully polarized state with $N$ spins up ($N$ spins down),  which no more spinons can be added to (removed from).

Expanding the spinon field about $\pm k_F$, $\Psi_s\sim e^{ik_Fx}\psi_{sR}+e^{-ik_Fx}\psi_{sL}$, we can write a phenomenological Hamiltonian density of the form\be
\mathcal{H}=\mathcal{H}_s+\mathcal{H}_c+\mathcal{H}_{cs},
\ee
where \bea
\mathcal{H}_s&=& \psi^\dagger_{sR}(-iv_s^{(0)}\partial_x+i\gamma^{(0)}\partial_x^3)\psi_{sR}^{\phantom\dagger}\nonumber\\
&&+ \psi^\dagger_{sL}(iv_s^{(0)}\partial_x-i\gamma^{(0)}\partial_x^3)\psi_{sL}^{\phantom\dagger}+\mathcal{H}_s^{int}
\eea
is the spinon Hamiltonian with spinon-spinon interactions contained in $\mathcal{H}_s^{int}$, $\mathcal{H}_c$ is the charge Hamiltonian given by the charge part of Eq. (\ref{eq:luttingermodel}), and $\mathcal{H}_{cs}$ contains spinon-holon interactions. Due to spinon particle-hole symmetry, the latter can only contain irrelevant operators (dimension three and higher), for instance, $ \psi^\dagger_{s,R/L}\partial_x\psi_{s,R/L}^{\phantom\dagger}\partial_x \varphi^c_{R/L}$.

The parameters $v_s^{(0)}$ and $\gamma^{(0)}$ are renormalized by interactions (both spinon-spinon and spinon-holon). We denote the parameters of the exact spinon dispersion by $v_s$ and $\gamma$. As usually done for $v_s$, the exact $\gamma$ can be extracted from the Bethe ansatz solution in the case of integrable models. For repulsive electron-electron interactions, we expect $v_s<v_c$, where $v_c$ is the exact charge velocity. In this case, the lower edge of the spectrum of $any$ dynamical correlation function at small momentum $q$ (such that $|\gamma q^2|\ll v_s<v_c$) is controlled by the spinon branch line, with a single ``deep spinon'' with energy $\epsilon_s(q)\equiv\epsilon_{sR}(q)\approx v_sq-\gamma q^3$ and a certain number of  spinon or holon excitations at the Fermi points.

Whether or not the band curvature of the spinon dispersion can be neglected in the calculation of dynamical quantities depends on the frequency range of interest.\cite{imambekov} Far enough from the threshold, for $|\omega-\epsilon_s(q)|\gg \gamma q^3$, we are allowed to drop the band curvature operators and the dynamics is captured by Luttinger liquid theory. We can  bosonize $\psi_{s,R/L}$ in the standard way,\cite{Giamarchi}  using  $\psi_{s,R/L} \sim  e^{-i \sqrt{2 \pi} \phi^s_{R/L}}/\sqrt{2\pi \alpha}$, with chiral bosonic fields $ \phi^s_{R/L}$ and $\alpha$ a short distance cutoff. The quadratic Hamiltonian in terms of spin bosons is diagonalized by a transformation to $\varphi^s_{R/L}$
\be \varphi^s_{L} \pm \varphi^s_{R} = K^{\pm \frac{1}{2}} (\phi^s_{L} \pm \phi^s_{R}) ,\ee
where $K$ is the Luttinger parameter for the spinons. The latter is fixed by SU(2) symmetry.
In Abelian bosonization, we write $S^z \sim \partial_x \varphi^s_{R} - \partial_x \varphi^s_{L}$, which has scaling dimension 1. For $S^+$, we bosonize the fermion and string operators in Eq.~(\ref{JW2}) and obtain 
\bea
S^+ \sim e^{- i \sqrt{2 \pi} \left(\sqrt{K}-\frac{1}{2 \sqrt{K}}\right) \varphi^s_{R}}  e^{ i \sqrt{2 \pi} \left(\sqrt{K}+\frac{1}{2 \sqrt{K}}\right) \varphi^s_{L}},
\eea
which has scaling dimension $K +1/4K$. Demanding that this dimension is equal to 1 gives  $K=1/2$. 

On the other hand, for $|\omega-\epsilon_s(q)|\ll \gamma q^3$, the behavior of dynamical correlation functions is sensitive to the band curvature energy scale. In this regime, it is important to consider that spinons created near the threshold with energy $\epsilon_s(q)$ travel with a different velocity than spinons at the spinon Fermi surface. It has become standard to treat this type of problem using quantum impurity models in analogy with the x-ray edge singularity.\cite{pustilnik06}
In order to describe the deep spinon threshold, we expand the spinon field with low and high energy subbands:\bea \label{d1d2}
\psi_{sR}&\sim& \psi_{sr} +e^{-iqx} d^\dagger_{s1}+e^{iqx} d^{\phantom\dagger}_{s2}.
\eea
The subband momentum cutoff is taken to be small compared to $q$. For $q>0$, we consider deep holes with momentum about $k_F-q$ and high-energy particles with momentum about $k_F+q$. These two types of excitations are degenerate as a consequence of particle-hole symmetry. The low energy   $\psi_{sr}$ and $\psi_{sl}$ fields are then bosonized, while $d_{s1,2}$ are treated as mobile impurities. We denote by $\varphi^s_{r/l}$ the chiral spin boson fields with the reduced cutoff at scale $q$. The interactions among low and high energy spinons and low energy charge bosons can be described by the effective Hamiltonian density\be
\mathcal{H}=\mathcal{H}_c+\mathcal{H}_{s,\ell}+\mathcal{H}_{d}+\mathcal{H}_{sd}+\mathcal{H}_{cd}.
\ee
Here $\mathcal{H}_c$ is the free charge boson Hamiltonian with reduced cutoff at scale $q$ (fields denoted by $\varphi^c_{r/l}$). In the spin-only part of $\mathcal H$,  \be
\mathcal{H}_{s,\ell}=\frac{v_s}{2}[(\partial_x \varphi^s_{r})^2 + (\partial_x \varphi^s_{l})^2]
\ee
is the Luttinger model for low energy spin excitations (here written in Abelian bosonization form), \be
\mathcal{H}_{d}=d^{\dagger}_{s1}[\epsilon_s(q)-iu\partial_x]d^{\phantom\dagger}_{s1}+(1\to2)
\ee
is the kinetic energy of the high energy spinons, with $u=u(q)$ the corresponding exact velocity, and\be
\mathcal{H}_{sd}=\frac{1}{\sqrt{4 \pi }}(V_l  \partial_x \varphi^s_{l} - V_r \partial_x \varphi^s_{r}) (d_{s2}^\dagger d_{s2}^{\phantom\dagger}-d_{s1}^\dagger d_{s1}^{\phantom\dagger} )
\ee
is the coupling between high energy spinons and low energy spin bosons. The amplitudes $V_r$ and $V_l$ scale $\sim q^2$ because they represent short-range interactions between spinless fermions and vanish as $q\to0$.\cite{pustilnik,pereiraPRL} Note that $\mathcal{H}_{sd}$ is invariant under the  particle-hole  transformation $\psi_{s,R/L}^{\phantom\dagger}\to \psi^\dagger_{s,R/L}$, which takes $\varphi^s_{r/l}\to-\varphi^s_{r/l}$, $d_{s1}^{\phantom\dagger}\leftrightarrow d_{s2}^{\phantom\dagger}$. Here we have neglected the backscattering operator $g$ in Eq. (\ref{eq:luttingermodel}), which becomes marginal at the SU(2) point; however, we point out that it might be important for logarithmic corrections to  edge singularities, which are known to exist for the DSSF of the Heisenberg spin chain.\cite{karbach} The spin-charge coupling in $\mathcal{H}$ is given by\be
\mathcal{H}_{cd}=\frac{1}{\sqrt{2 \pi K_c}}(V_l^c \partial_x \varphi^c_{l} -V_r^c \partial_x \varphi^c_{r}) (d_{s1}^\dagger d_{s1}^{\phantom\dagger}+d_{s2}^\dagger d_{s2}^{\phantom\dagger}). \label{Hcsimp}
\ee
The amplitudes $V_{r/l}^c$ in Eq. (\ref{Hcsimp}) stem from operators like $ \psi^\dagger_{sR}\partial_x\psi_{sR}^{\phantom\dagger}\partial_x \varphi^c_{r/l}$, taking the high-energy mode in the expansion of $\psi_R$. As a result, $V_{r/l}^c$ scale like $\sim q$. The relation to the parameters in Eq. (\ref{xray}) is $V^c_{l/r}=\pm\sqrt{2K_c}\pi q\kappa_\pm^\prime$.

Hamiltonian $\mathcal{H}$ can be diagonalized by a unitary transformation of the form $\tilde H = U H U^\dagger$ where $U=U_1U_2$ with \bea 
U_{1,2}=e^{ -i \int dx \left(\mp\frac{\gamma_r \varphi^s_{r} + \gamma_l \varphi^s_{l}}{\sqrt{ \pi }}+\frac{\gamma^c_r \varphi^c_{r} + \gamma^c_l \varphi^c_{l}}{\sqrt{2 \pi K_c}}  \right) d_{s1,2}^\dagger d_{s1,2}^{\phantom\dagger}},
 \eea
with \be 
\label{lambdakappa} \gamma_{l/r} =- \frac{V_{l/r}}{2(v_s \pm u)},~~~ \gamma_{l/r}^c = -\frac{V^c_{l/r}}{v_c \pm u}\approx- \frac{V^c_{l/r}}{v_c \pm v_s}.  \ee
This transformation takes 
\be
d_{s1,2} \to \tilde{d}_{s1,2} e^{ -i  \left(\mp\frac{\gamma_r \varphi^s_{r} + \gamma_l \varphi^s_{l}}{\sqrt{\pi}}+\frac{\gamma^c_r \varphi^c_{r} + \gamma^c_l \varphi^c_{l}}{\sqrt{2 \pi K_c}}  \right) },
\ee 
where $ \tilde{d}_{s1,2} $ are free (up to irrelevant operators). The $\gamma$'s are interpreted as phase shifts at  the spinon and holon Fermi points due to the creation of a high-energy spinon.

The exponents for the edge singularities are then calculated using the methods of Refs. \onlinecite{pustilnik06,pereiraPRL}.
The threshold for the longitudinal DSSF $S^{zz}(q,\omega)$ is given by the correlation function for the operator that creates a particle-hole pair of spinons with a hole at $k_F$ and a particle at $k_F+q$ (or equivalently a hole at $k_F-q$ and a particle at $k_F$) 
\be \label{B0}  B_{z}^{\dagger} = d_{s2}^\dagger \psi_{sr}\sim \tilde{d}_{s2}^\dagger e^{-i \sqrt{2 \pi} (\lambda_r \varphi^s_{r}+\lambda_l \varphi^s_{l}+\lambda_r^c \varphi^c_{r}+\lambda_l^c \varphi^c_{l})},  \ee
with
\bea
\lambda_r &=&\frac{1}{\sqrt{2}} \left( \frac{3}{2} -\frac{\gamma_r}{\pi}  \right), \nonumber \\
\lambda_l &=&\frac{1}{\sqrt{2}} \left( \frac{1}{2} -\frac{\gamma_l}{\pi}  \right), \eea 
and \be \label{lambdaRLc}
\lambda_{r/l}^c =-\frac{1}{2 \sqrt{K_c}} \frac{\gamma_{r/l}^c}{\pi}. \ee
Using Eq.~(\ref{B0}), we calculate  the Fourier transform of the correlation function $\langle B_z(x,t) B^\dagger_z(0,0)\rangle$  and find a power-law singularity  $S^{zz}(q, \omega) \sim(\omega - \omega_{s-})^{\mu_{zz}}$ with exponent
\bea
\label{muzz}
\mu_{zz} &=&-1 + (\lambda_r)^2 + (\lambda_l)^2 + (\lambda_r^c)^2 + (\lambda_l^c)^2  \nonumber \\
&=&-1+\frac{1}{2}\left( \frac{3}{2}-\frac{\gamma_r}{\pi}\right)^2 + \frac{1}{2}\left( \frac{1}{2}-\frac{\gamma_l}{\pi} \right)^2 \nonumber \\
& &+\frac{1}{4K_c}\left[\left(\frac{\gamma_r^c}{\pi}\right)^2+\left(\frac{\gamma_l^c}{\pi}\right)^2\right] .\eea
The last term amounts to an orthogonality catastrophe contribution to the exponent due to coupling of the   $d_{s1,2}$ particles to gapless charge modes. 

Now  consider the transverse DSSF $S^{+-}(q,\omega)$. In this case the operator Eq.~(\ref{JW2}) creates a particle and a string 
\bea \label{C0} B_{+}^\dagger(x) & \sim& d_{s2}^\dagger e^{-i \sqrt{\pi/2}(\phi^s_l-\phi^s_r)} \nonumber \\ &\sim&  \tilde{d}_{s2}^\dagger e^{-i \sqrt{2 \pi} (\lambda_r^\prime \varphi^s_{r}+\lambda_l^\prime\varphi^s_{l}+{\lambda_r^c}\varphi^c_{r}+{\lambda_l^c}\varphi^c_{l})}  \eea 
with  \bea 
\label{lambdap} \lambda_{r}^\prime =\frac{1}{\sqrt{2}} \left(\frac{1}{2} - \frac{\gamma_{r}}{\pi} \right), ~~~\  \lambda_{l}' =\frac{1}{\sqrt{2}} \left(-\frac{1}{2} - \frac{\gamma_{l}}{\pi} \right).
\eea 
The Fourier transform of the correlation function $\langle B_+(x,t)B^\dagger_+(0,0) \rangle$ leads to $S^{+-}(q,\omega) \sim (\omega - \omega_{s-})^{\mu_{+-}}$ with the exponent
\bea
 \mu_{+-}&=& -1 + \frac{1}{2} \left( \frac{1}{2} - \frac{\gamma_r}{\pi} \right)^2+ \frac{1}{2} \left( \frac{1}{2} + \frac{\gamma_l}{\pi} \right)^2  \nonumber \\
& &+\frac{1}{4K_c}\left[\left(\frac{\gamma_r^c}{\pi}\right)^2+\left(\frac{\gamma_l^c}{\pi}\right)^2\right] . \eea
SU(2) symmetry implies that $\mu_{zz}=\mu_{+-}$, but this is only one equation. However, we can actually get infinitely many equations by imposing SU(2) symmetry for the  singularities that differ from the above by zero-energy excitations in which spinons are transferred between the Fermi points. These are created by electron backscattering processes in the language of spin-1/2 fermions, which are umklapp processes for spinons,
\be (\psi_{sr}^\dagger \partial_x \psi_{sr}^\dagger  \psi_{sl} \partial_x \psi_{sl})^n \sim e^{-i \sqrt{4 \pi} n (\varphi^s_{l} -  \varphi^s_{r})} ,\label{backn}\ee
where $n$ is an integer ($n<0$ on the right-hand side of Eq. (\ref{backn}) corresponds to the Hermitian conjugate of the left-hand side).  The excitations that differ in momentum by $2nk_F$ have thresholds at the same frequency $\omega_{s-}(q)$ because the spin spectrum is periodic in momentum with period $2k_F$. The exponents for $|q-2nk_F|\ll k_F$ are given by
\bea
\mu_{zz,n}  &=& -1 + (\lambda_{r,n})^2 + (\lambda_{l,n})^2 + (\lambda_r^c)^2 + (\lambda_l^c)^2   ,\nonumber\\
\mu_{+-,n}  &=& -1 + (\lambda_{r,n}^\prime)^2 + (\lambda_{l,n}^\prime)^2 + (\lambda_r^c)^2 + (\lambda_l^c)^2  ,\nonumber
\eea
where \bea
\lambda_{r,n} &=&\frac{1}{\sqrt{2}} \left( -2n+\frac{3}{2} -\frac{\gamma_r}{\pi}  \right), \nonumber \\
\lambda_{l,n} &=&\frac{1}{\sqrt{2}} \left( +2n+\frac{1}{2} -\frac{\gamma_l}{\pi}  \right), \\
\lambda_{r,n}^\prime &=&\frac{1}{\sqrt{2}} \left( -2n+\frac{1}{2} -\frac{\gamma_r}{\pi}  \right), \nonumber \\
\lambda_{l,n} ^\prime&=&\frac{1}{\sqrt{2}} \left( +2n-\frac{1}{2} -\frac{\gamma_l}{\pi}  \right).\nonumber
\eea

The condition $\mu_{zz,n} =\mu_{+-,n} $ is satisfied for all $n$ if and only if $\gamma_r/\pi=\gamma_l/\pi=1/2$. With this result, the  exponent in the lower edge of the DSSFs (or the DCSF) for $q\ll k_F$ becomes \be
\mu_{s-}=\mu_{zz}=\mu_{+-}=-\frac12+\frac{1}{4K_c}\left[\left(\frac{\gamma_r^c}{\pi}\right)^2+\left(\frac{\gamma_l^c}{\pi}\right)^2\right] .
\ee
With $\gamma_{l/r}^c$ given in Eq. (\ref{lambdakappa}) and $V_{l/r}^c=\pm\sqrt{2K_c}\pi q\kappa_\pm^\prime$ fixed as explained in section \ref{sec:spinedge}, we obtain the final result in Eq. (\ref{mu}).


\begin{thebibliography}{12}
\bibitem{comment1}The $\sqrt{2}$ factor was missed in Ref. \onlinecite{teber}.
\bibitem{comment} For $\nu_1=\nu_2$ and low enough temperatures, the drag would be dominated by backscattering ($q\sim 2k_F$) processes; see Refs. \onlinecite{nazarov} and \onlinecite{pustilnik}.
\end{thebibliography}
\end{document}